\documentclass[11pt]{article}
\usepackage{aaspp4}
\newcommand{\delnud}{\Delta\nu_{\mathrm{d}}}
\newcommand{\ths}{\theta_{\mathrm{s}}}

\newcommand{\thi}{\theta_{\mathrm{i}}}
\newcommand{\thapp}{\theta_{\mathrm{app}}}

\newcommand{\cnsq}{C_n^2}

\newcommand{\smu}{\mbox{kpc~m${}^{-20/3}$}}

\newcommand{\sech}{\mathop{\mathrm{sech}}\nolimits}
\newcommand{\erf}{\mathop{\mathrm{erf}}\nolimits}

\newcommand{\hin}{h_{1,\mathrm{in}}}
\newcommand{\hout}{h_{1,\mathrm{out}}}
\newcommand{\fin}{F_{1,\mathrm{in}}}
\newcommand{\fout}{F_{1,\mathrm{out}}}

\newcommand{\mion}[2]{\mathrm{#1}\:\mathrm{\expandafter\uppercase\expandafter{\romannumeral#2}\relax}}

\lefthead{Lazio \& Cordes}
\righthead{Angular Broadening Toward the Anticenter}

\received{1997 June 5}
\accepted{1997 November 14}
\paperid{MS~36593}

\begin{document}

\title{The Radial Extent and Warp of the Ionized Galactic Disk.  II.\\
	A Likelihood Analysis of Radio-Wave Scattering \\
	Toward the Anticenter}

\author{T.~Joseph W.~Lazio\altaffilmark{1,2} \& James M.~Cordes}
\affil{Department of Astronomy and National Astronomy \& Ionosphere
	Center, Cornell University, Ithaca, NY  14853-6801; \\
	lazio@spacenet.tn.cornell.edu, cordes@spacenet.tn.cornell.edu}
\authoraddr{T. Joseph W. Lazio
            NRL, Code 7210
            Washington, DC  20375-5351}
\altaffiltext{1}{NRC/NRL Post-doctoral Associate}
\altaffiltext{2}{Current address: NRL, Code~7210, Washington, DC  20375-5351;
	lazio@rsd.nrl.navy.mil}

\begin{abstract}
We use radio-wave scattering data for extragalactic sources and
pulsars to constrain the distribution of ionized gas in the outer
Galaxy.  Like previous models, our model for the \ion{H}{2} disk
includes parameters for the radial scale length and scale height of
the ionized gas.  In addition, we have used the known \ion{H}{1}
distribution in the outer Galaxy in constructing our model, and we
allow the \ion{H}{2} disk to warp and flare.  We also include the
Perseus arm in our model.  We use a likelihood analysis on 18
anticenter sources with measured scattering observables: 11
extragalactic sources and 7 pulsars.  We find that the strength of
scattering in the Perseus arm is no more than 60\% of the level
contributed by spiral arms in the inner Galaxy and is equivalent to a
scattering diameter of 1.5~mas at~1~GHz.  Our analysis favors an
unwarped, non-flaring disk with a scale height of 1~kpc, though this
may reflect the non-uniform and coarse coverage of the anticenter
provided by the available data.  One extragalactic source has a size a
factor of two smaller than predicted by our model, possibly indicating
the existence of holes in the scattering material.  The lack of a warp
in the scattering material indicates that VLBI observations near 1~GHz
with an orbiting station having baseline lengths of a few Earth
diameters will not be affected by interstellar scattering at moderate
Galactic latitudes, $|b| \approx 15\arcdeg$.  The radial scale length
is 15--20~kpc, but the data cannot distinguish between a gradual
decrease in the electron density and a truncated distribution.  We
favor a truncated one, because we associate the scattering with
massive star formation, which is also truncated near 20~kpc.  A radial
extent of 20~kpc is also comparable to the radial extent of H$\alpha$
emission observed for nearby spiral galaxies.  We find that the
distribution of electron density turbulence must decrease more rapidly
with Galactocentric distance than the distribution of hydrogen.
Alternate ionizing and turbulent agents---the intergalactic ionizing
flux and the passage of satellite galaxies through the disk---are
unlikely to contribute significant amounts to scattering in the
anticenter.  We cannot exclude the possibility that a largely ionized,
but quiescent disk extends to $\gtrsim 100$~kpc, similar to that
inferred for some Ly$\alpha$ absorbers.
\end{abstract}


\section{Introduction}\label{sec:intro}

Early investigations of the Galaxy's \ion{H}{1} emission revealed that
it extends well past the solar circle, and that, in the outer Galaxy,
the emission is warped systematically from its midplane in the inner
Galaxy (\cite{b57}; \cite{k57}; \cite{w57}; Oort, Kerr, \&
Westerhout~1958).  More recent stellar (\cite{ds89}), infrared
(\cite{sdhk87}; \cite{freudenreichetal94}), and molecular
(\cite{wbbk90}) observations have shown that these disk constituents
also extend well past the solar circle and are warped similar to the
\ion{H}{1} layer.

Ionized gas occupies potentially 10\% or more of the volume of the
interstellar medium (ISM) near the solar circle and is probably a
dynamically important constituent (e.g., \cite{r77}; \cite{kh87}), but
its radial extent is poorly constrained.  H$\alpha$ measurements are
limited to distances of a few kiloparsecs by interstellar absorption
(\cite{r83}).  The frequency at which the Galactic plane becomes
optically thick because of free-free absorption can indicate the
extent of the disk, but for plausible disk sizes (see below) this
frequency is less than 10~MHz and so is difficult to observe.  Few
pulsars are known in the anticenter direction.  Less than ten have
dispersion-measure--independent distance estimates ($\mathrm{DM} =
\int n_{\mathrm{e}}\,ds$) and the estimated distances are less than
2~kpc (\cite{fw90}); the remainder have DMs of 30--125~pc~cm${}^{-3}$,
consistent with distances of a few kiloparsecs (Taylor, Manchester, \&
Lyne~1993; \cite{zcwl96}).  Fluctuations in the ionized gas produce
radio-wave scattering which manifests itself as angular broadening of
compact sources (see \cite{r90} for a review of the full variety of
interstellar radio wave propagation effects).  Scattering measurements
have been biased toward the inner Galaxy, even for surveys of angular
broadening of extragalactic sources (e.g., Fey, Spangler, \&
Mutel~1989; Fey, Spangler, \& Cordes~1991).  Only one angular
broadening survey has been conducted toward the outer Galaxy
(\cite{dtbbbc84}), and, as we illustrate below, most of the sources in
that survey had Galactic latitudes too large to provide effective
constraints on the radial extent of the ionized Galactic disk.
Measurements of interplanetary scintillation determine source
diameters indirectly, but in general do not have sufficient resolution
to provide stringent constraints.

Though the radial extent of the ionized gas is poorly constrained, a
number of lines of evidence suggest that its radial extent may equal
or exceed that of the \ion{H}{1}:
\begin{itemize}
\item Savage, Sembach, \& Lu~(1995) find \ion{C}{4} absorption along
the line of sight to H~1821$+$643 ($\ell = 94\arcdeg, b=27\arcdeg$).
Among the velocity components contributing to this absorption is
low-density ($n \sim 5.6 \times 10^{-3}$~cm${}^{-3}$), warm ($T \sim
10^4$~K) gas at a velocity of $-120$~km~s${}^{-1}$, corresponding to a
kinematic Galactocentric distance of 25~kpc.

\item The \ion{H}{1} disks of nearby galaxies are truncated at radii of
order 25--50~kpc, at which the surface density drops to
$N_{\mion{H}{1}} \lesssim 2 \times 10^{19}$~cm${}^{-2}$ (Corbelli,
Schneider, \& Salpeter~1989; \cite{v91}; \cite{b-h97}).  This
truncation is observed even for galaxies without nearby companions and
likely occurs where the disks become optically thin to the
intergalactic ionizing flux (\cite{s69}; \cite{ecees93}).
Bland-Hawthorn, Freeman, \& Quinn~(1997) have reported the detection
of ionized gas beyond the observed \ion{H}{1} disk of NGC~253.
Charlton, Salpeter, \& Hogan~(1992) have proposed that at least some
of the low-redshift Ly$\alpha$ clouds seen in quasar spectra may be
due to residual \ion{H}{1} in extended, nearly fully ionized disks of
normal spiral galaxies.  Our Galaxy would then be a prototypical, $z =
0$ absorber.

\item Material blown out of the Galactic
disk by the action of clustered supernovae may account for a fraction
of high-velocity clouds and later return to the disk forming a
Galactic fountain (\cite{sf76}; \cite{b80}; \cite{hb90}; \cite{s90};
\cite{k91}).  Models of high-velocity clouds often require the
material to be supported by gas pressure at large Galactocentric
radii, $R \gtrsim 25$~kpc (e.g., \cite{b80}).
\end{itemize}

Taylor \& Cordes~(1993, hereinafter \cite{tc93}) modelled the Galactic
distribution of ionized gas with three global components:
\begin{enumerate}
\item an extended component with scale height of approximately 1~kpc and
$1/e$ radial scale length of order 20~kpc;
\item an inner Galaxy component with scale height of 0.15~kpc and
radial scale length of 3.5~kpc; and
\item spiral arms, the number and shape of which were determined by appeal to
radio and optical observations of \ion{H}{2} regions and radio
observations of \ion{H}{1} and non-thermal emission.
\end{enumerate}
The data available to constrain the model parameters consisted of 74
pulsars with DM-independent distances, 223 scattering measurements
toward pulsars, masers, and extragalactic sources, and the Galactic
longitude distribution for 553 pulsar DMs.

Because of the paucity of measurements summarized above, \cite{tc93}
could place only a lower bound on the scale length of the extended
component, $A_1$.  They adopted $A_1 = 20$~kpc, though $A_1 \approx
50$~kpc produced comparable fits to the data.  Moreover, they modelled
the extended component as planar; if the ionized disk does extend to
20~kpc or more, it is likely to be warped similarly to the other outer
Galaxy disk constituents.

Figure~\ref{fig:angle} demonstrates that angular broadening
measurements of extremely low-latitude, $|b| < 1\arcdeg$,
extragalactic sources toward the Galactic anticenter have the
potential of constraining $A_1$.  The line of nodes of the \ion{H}{1}
disk is fairly constant with Galactocentric radius and is near a
Galactic longitude of 170\arcdeg, so sources toward the anticenter
probably provide the longest path length through the ionized disk.
Extremely low-latitude sources are required because the scale height
of the extended component near the solar circle is 0.88~kpc and only
for sources with $|b| < 1\arcdeg$ does the line of sight remain within
one scale height for path lengths of 50~kpc or more.  Only three of
the sources in Dennison et al.'s~(1984) survey meet this criterion; of
these, one may show excessive scattering due to an \ion{H}{2} region
along the line of sight and another shows complex structure making it
difficult to estimate a scattering diameter.

We have carried out a program of multifrequency Very Long Baseline
Array (VLBA) observations of twelve anticenter sources, seven of which
have $|b| < 0\fdg5$ (Lazio \& Cordes~1997, hereinafter \cite{lc97}).
We detected all but one of the sources at one or more of our
observation frequencies---0.3, 1.6, and 5~GHz.  As
Fig.~\ref{fig:angle} illustrates, the nominal resolutions of the VLBA
are such that 18~cm observations are sensitive to scale lengths of
$A_1 \gtrsim 100$~kpc and~90~cm observations should detect scattering
even if $A_1 \lesssim 10$~kpc.

Figure~\ref{fig:angle} also shows the nominal resolution of the space
VLBI satellite HALCA, and thereby illustrates another important aspect
of improving our knowledge of the Galactic distribution of scattering.
At low Galactic latitudes, interstellar scattering will determine the
limiting resolution for baselines in excess of the Earth's diameter at
frequencies near 1~GHz.  If the \ion{H}{2} disk flares or warps
similarly to the \ion{H}{1} disk, interstellar angular broadening
could be non-negligible at much higher latitudes (e.g., $|b| \approx
30\arcdeg$).

In this paper we combine the sources from our survey with
other radio-wave scattering measurements from the literature and use a
likelihood analysis to constrain the distribution of ionized gas in
the outer Galaxy.  In \S\ref{sec:outerdisk} we describe our model for
the distribution of ionized gas in the outer Galaxy, in
\S\ref{sec:analyze} we extract scattering diameters from our measured
angular diameters and develop a likelihood analysis of scattering in
the outer Galaxy, and in \S\ref{sec:conclude} we discuss our results
and present our conclusions.

\section{A Model of the Ionized Disk in the Outer Galaxy}
\label{sec:outerdisk}

In this section we develop a model for the distribution of free
electrons in the outer Galaxy.  Based on the close correspondence
between the \ion{H}{1} disk and other outer Galaxy constituents,
\S\ref{sec:intro}, we shall use the \ion{H}{1} distribution in the
outer Galaxy as a basis for modifying the \cite{tc93} model.  We begin
with a discussion of the connection between the observed scattering
angle, $\ths$, and the (modelled) electron density, $n_{\mathrm{e}}$.

\subsection{Electron Density Fluctuations and Angular Broadening}

The density fluctuations responsible for angular broadening are
parameterized commonly with a power-law spectrum, 
\begin{equation}
P_{\delta n} = \cnsq q^{-\alpha}, 
\label{eqn:spectrum}
\end{equation}
over a range of spatial wavenumbers, $q_0 \ll q \ll q_1$, where $l_0 =
2\pi/q_0$ and $l_1 = 2\pi/q_1$ are the outer and inner scales, respectively,
to the spectrum.  The quantity $\cnsq$ sets the amplitude of the density
fluctuations and varies spatially.  Throughout we adopt a spectral index of
$\alpha = 11/3$, the Kolmogorov value, as suggested by a number of
observations (\cite{r90}).

The scattering angle for plane-wave radiation propagating a
distance~$D$ through a medium filled with such a spectrum of density
fluctuations is (\cite{cwfsr91}; \cite{cl91})
\begin{equation}
\ths = 128\,\mathrm{mas}\,\mathrm{SM}^{3/5}\nu_{\mathrm{GHz}}^{-11/5}.
\label{eqn:scattangle}
\end{equation}
The quantity SM is the line-of-sight integral of $\cnsq$, 
\begin{equation}
\mathrm{SM} = \int_0^D ds\,\cnsq(s),
\end{equation}
and $\nu_{\mathrm{GHz}}$ is the frequency in GHz.

Cordes et al.~(1991) demonstrated that the level of scattering, as
measured by SM, correlates with the dispersion measure, DM, for nearby
pulsars; for pulsars toward the inner Galaxy, the level of scattering
increases faster with distance than \hbox{DM}.  This correlation suggests
that the electrons responsible for dispersion are also responsible for
scattering.  Cordes et al.~(1991) and \cite{tc93} adopted
\begin{equation}
d\mathrm{SM} = C_{\mathrm{u}}Fn_{\mathrm{e}}^2\,ds.
\label{eqn:sm}
\end{equation}
Here $n_{\mathrm{e}}$ is the electron density in cm${}^{-3}$, $F$ is
the fluctuation parameter and is a
measure of how effectively density fluctuations are produced and
maintained, $ds$ is a path length interval in kpc, and
$C_{\mathrm{u}}$ is a constant responsible for producing SM in the
typical units of \smu, $C_{\mathrm{u}} =
3.4(2\pi)^{-1/3}$~m${}^{-20/3}$~cm${}^6$.  With a model for
$n_{\mathrm{e}}$ in the outer Galaxy, we can integrate
equation~(\ref{eqn:sm}) along the line of sight to a source to find SM
and $\ths$.

\subsection{Free Electron Density Model}\label{sec:ac.model}

Of the four components in the \cite{tc93} model, only the spiral arms
and the extended component are relevant to our study of the outer
Galaxy.  The inner Galaxy component has a scale length of 3.5~kpc and,
at the solar circle, its contribution to the electron density has
decreased to 0.2\% of that from the extended component.  The Gum
Nebula, which was also included in the model because of its proximity
to the Sun, only affects lines of sight within about 20\arcdeg\ of
$(\ell, b) = (260\arcdeg, 0\arcdeg)$, well removed from the directions
to the sources considered here.

We retain the spiral arm component because one of the spiral arms, the
Perseus arm, is outside the solar circle over the longitude range of
interest.  Lines of sight with $|b| \lesssim 10\arcdeg$ pass within
one scale height of the center of this spiral arm.  In the \cite{tc93}
model, this arm contributes a scattering measure of $\mathrm{SM} \sim
0.01$~\smu, equivalent to a 1~GHz scattering angle of $\ths \sim
10$~mas for the line of sight $(\ell, b) = (180\arcdeg, 0\arcdeg)$.

The remaining component is the extended component.  In the \cite{tc93}
model this component consisted of a flat disk, centered on the
Galaxy's midplane with a scale height $h_1 = 0.88$~kpc and scale
length $A_1 = 20$~kpc.  We use \ion{H}{1} observations toward the
outer Galaxy as a guide for constructing our model for three reasons.
First, in the inner Galaxy, the mean and rms electron density
distributions generally follow the distribution of massive stars, and,
toward the outer Galaxy, sites of massive star formation follow the
\ion{H}{1} distribution (\cite{wbbk90}).  Second, Savage et al.~(1995)
detected ionized gas that is spatially coincident with warped
\ion{H}{1} gas.  Finally, models of low-redshift Ly$\alpha$ clouds, in
which the outer extent of the Galaxy is nearly fully ionized by the
intergalactic ionizing flux (\cite{csh93}), predict that the
\ion{H}{1} disk is ionized to form the \ion{H}{2} disk.  The
\ion{H}{1} structure in the outer disk has been reviewed by
Burton~(1992).  Here we shall only summarize salient details as we
construct our model for the outer \ion{H}{2} disk.

In the outer Galaxy the distribution of $n_{\mathrm{e}}$ is
the sum of the extended component and the Perseus arm,
\begin{equation}
n_{\mathrm{e}}(x, y, z) = n_1g_1(R)\sech^2[Z(R)/h_1(R)]
 + f_4n_{\mathrm{a}}\sech^2(z/h_{\mathrm{a}})g_{\mathrm{a}}(R, d).
\label{eqn:ne}
\end{equation}
The nominal densities of the two components are $n_{1,\mathrm{a}}$;
their radial dependences are given by the functions
$g_{1,\mathrm{a}}$; and their scale heights by $h_{1,\mathrm{a}}$,
respectively.  The midplane of the extended component is given by
$Z(R)$ and the Perseus arm has a fine-tuning parameter of $f_4$.  The
following sections explain the various quantities in more detail.  As
in \cite{tc93} the coordinate system has the $x$-axis directed
parallel to $\ell = 90\arcdeg$, the $y$-axis directed parallel to
$\ell = 180\arcdeg$, the $z$-axis directed toward $b > 0\arcdeg$, and
the Galactocentric radius is $R = \sqrt{x^2 + y^2}$.  The minimum
distance between the position $(x, y)$ and a point on the Perseus arm
is denoted by $d$ (see \cite{tc93} for a full description of the
spiral arms).  Following \cite{tc93} we take the radial and $z$
dependences to be separable.

\subsubsection{Radial Dependence of the Extended Component}\label{sec:ac.radial}

We consider two functional forms for the radial dependence of the
extended component, $g_1(R)$.  The first is a $\sech^2$ dependence
\begin{equation}
g_1^{(1)}(R)
 = \sech^2\left(R/A_1^{(1)}\right)\big/\sech^2\left(8.5\,\mathrm{kpc}/A_1^{(1)}\right).
\label{eqn:radial1}
\end{equation}
We shall refer to this as a $\sech^2$ disk.  This functional form
exhibits a gradual decrease in the electron density with $R$; for $R
\gg A_1^{(1)}$, $g_1^{(1)}(R) \propto \exp\left(-2R/A_1^{(1)}\right)$.
The second form is
\begin{equation}
g_1^{(2)}(R) = \left\{\begin{array}{ll}

	\cos\left(\pi R/2A_1^{(2)}\right)\big/\cos\left(\pi R_{\sun}/2A_1^{(2)}\right), & \mbox{$R \le A_1^{(2)}$}; \\
	0, & \mbox{$R > A_1^{(2)}$}.
	
		      \end{array}
\right .
\label{eqn:radial2}
\end{equation}
We shall refer to this as
a truncated disk because it is zero for $R > A_1^{(2)}$.

Figure~\ref{fig:radial} compares these functional forms with each
other and with the \ion{H}{1} density (\cite{gb76}).  Our choice for
these particular functional forms is motivated by a number of
considerations: The $\sech^2$ dependence is that assumed by \cite{tc93} and
so our results can be compared directly to theirs.  When compared to
the \ion{H}{1} density, the $\sech^2$ disk has a slower radial fall off
while the truncated disk has a faster radial fall off.  These two
functional forms should bracket the actual $A_1$ if the electron
density distribution follows that of the \ion{H}{1}.  The truncated
disk also allows us to model a disk with variable scattering
properties.  At $R > A_1^{(2)}$ there is no additional scattering.  As
we have written equation~(\ref{eqn:radial2}), this truncation occurs
because $n_{\mathrm{e}} = 0$~cm${}^{-3}$ for $R > A_1^{(2)}$.  An
equivalent model is one in which there is ionized gas but $F_1 = 0$
for $R > A_1^{(2)}$.  Such a truncation could occur if the
distribution of scattering agents decreased more rapidly with $R$ than
does $n_{\mathrm{e}}$.  Finally, our estimates of $A_1$ are model
dependent.  Comparison of the two forms allows us to assess the
sensitivity of our estimates of $A_1$ to the assumed models.  Both
functions are normalized so that $g_1(R_{\sun}) = 1$.  Henceforth, we
shall drop the superscripts (1) and (2) as it will be clear from the
context which $A_1$ parameter we are describing.

\subsubsection{$z$-dependence of the Extended Component and the Galactic Warp}\label{ac.z}

The $z$-dependence of the outer Galaxy density distribution is allowed
to differ from that of the \cite{tc93} model by two effects.  First,
the \ion{H}{1} disk is observed to flare to larger scale heights as
$R$ increases.  We model this effect by allowing the scale height to
vary with Galactocentric distance, $h_1(R)$.  The second is that the
\ion{H}{1} layer is warped.  The actual shape of the \ion{H}{1} warp
is complex, with differences between the northern and southern
Galactic hemispheres and with an amplitude that is radially dependent
(though other tracers of the outer disk are more symmetric than the
\ion{H}{1}, \cite{ds89}; \cite{wbbk90}).  Because the longitude range
of our data, $150\arcdeg < \ell < 210\arcdeg$, is significantly less
than the 180\arcdeg\ longitude range over which the north/south
asymmetry is important, we shall ignore the asymmetry in the warp.

The scattering measure is an integrated quantity, while \ion{H}{1} and
CO measurements yield velocity information.  We therefore model the
outer ionized disk as a single tilted ring or torus.  Our
approximation to the disk in the outer Galaxy is illustrated in
Fig.~\ref{fig:ac.geometry}.

The \ion{H}{1} disk begins to warp and flare significantly at the same
radius, $R \approx 10.5$~kpc.  The line of nodes of the \ion{H}{1}
disk varies with $R$, but is approximately centered on $\ell =
170\arcdeg$.  The onset of the \ion{H}{2} warp is assumed to occur at
the same radius as the \ion{H}{1} warp does.  At $R < 10.5$~kpc, our
model agrees with the \cite{tc93} model, with a scale height, $\hin$,
and fluctuation parameter, $\fin$.  At $R > 10.5$~kpc the \ion{H}{2}
disk is tilted by an angle $\Psi$ with respect to $b = 0\arcdeg$, has
a line of nodes $\ell_0^\prime$, a scale height $\hout$, and
fluctuation parameter $\fout$.  The radial dependences in
equations~(\ref{eqn:radial1}) and (\ref{eqn:radial2}) are continuous
through $R = 10.5$~kpc.

In the warped disk, the $z$-dependence and scale height in
equation~(\ref{eqn:ne}) are relative to the midplane of the gas,
\emph{not} to $b = 0\arcdeg$, cf.\ Fig.~\ref{fig:ac.geometry}.
Interior to the warp, the distance above the plane is $Z = s\sin b$
where $s$ is a distance along the line of sight.  Within the warp a
source with with Galactic coordinates $(\ell, b)$ has a latitude
$b^\prime$ relative to the torus' midplane.  The corresponding
$z$-height is $Z = s\sin b^\prime$ where $b^\prime$ is given by 
\begin{equation}
\sin b^\prime
 = \sin b\cos\Psi + \cos b\sin(\ell - \ell_0^\prime)\sin\Psi.
\label{eqn:bprime}
\end{equation}
The warped portion of the disk is assumed axisymmetric, as is the
unwarped extended component.

The division between $\fin$ and $\fout$ is to allow for the
possibility that the mechanism for generating or maintaining density
fluctuations in the far outer Galaxy may differ from that in the inner
Galaxy.  The radius at which the transition from $\fin$ to $\fout$
occurs was chosen to be that at which the disk begins to flare and
warp significantly.  The key assumption utilized here is that the
flaring and warping may be symptomatic of other processes which could
result in a change $\fin$ to $\fout$.

\subsubsection{Spiral Arm Component}

The functional dependences for the spiral arm component are unaltered
from the \cite{tc93} model.  In particular, the radial dependence for
the spiral arms is
\begin{equation}
g_{\mathrm{a}}(R, d) = \left\{
 \begin{array}{ll}
  \exp[-(d/w_{\mathrm{a}})^2], & R \le A_{\mathrm{a}}, \\
  \exp[-(d/w_{\mathrm{a}})^2]\sech^2[(R-A_{\mathrm{a}})/2], & R > A_{\mathrm{a}},
 \end{array}
 \right.
\end{equation}
with $A_{\mathrm{a}} = 8.5$~kpc a scale length analogous to $A_1$ and
$w_{\mathrm{a}} = 0.3$~kpc the width of a spiral arm.

\subsection{Preliminary Constraints on Model Parameters}

The source 87GB~0558$+$2325 has a measured diameter of approximately
4~mas at 1~GHz (\cite{lc97}), a factor of two less than that predicted
by the Perseus arm's contribution alone (\cite{tc93}).  In order that
our model not overpredict scattering diameters, we must modify the
spiral arm component as well.  Of the nine parameters\footnote{
Strictly speaking, the shapes of the arms are described by fiducial points
which are also parameters.  However, they have been determined largely from
radio and optical observations of \ion{H}{2} regions and radio
observations of thermal and \ion{H}{1} emission.  We have not altered
the fiducial points.}
describing the spiral arms---$n_{\mathrm{a}}$, $h_{\mathrm{a}}$,
$F_{\mathrm{a}}$, $w_{\mathrm{a}}$, $A_{\mathrm{a}}$, and four
fine-tuning parameters, $f_j$---\cite{tc93} used DM and scattering data to
fit for the first three.  They appealed to other radio and optical
observations in fixing $w_{\mathrm{a}} = 0.3$~kpc and $A_{\mathrm{a}}
= 8.5$~kpc.  In the inner Galaxy \cite{tc93} used the $f_j$ to obtain better
agreement between the model and observations toward the tangent points
of certain spiral arms.  In the \cite{tc93} model $f_4$, the
fine-tuning parameter for the Perseus arm, was set to unity; we shall
allow it to vary.

Our justification for allowing only $f_4$ to vary is that it is the
only spiral arm parameter that can be modified without affecting the
model in the inner Galaxy.  While the nine sources from our survey in
\cite{lc97} are a substantial fraction of the number of available
scattering measurements toward the anticenter, cf.\ 
Table~\ref{tab:literature} and Fig.~\ref{fig:vlbalocs}, the total
number of scattering measurements toward the anticenter is only
approximately twenty.  This number is a small fraction of the nearly
300 DM and SM measurements that \cite{tc93} used to constrain the model
parameters.  Thus, incorporating our additional measurements from
\cite{lc97} into the 300 used by \cite{tc93} and repeating their analysis would
not lead to any substantial change of the model in the inner Galaxy.

With the above model we can now integrate $d\mathrm{SM}$,
equation~(\ref{eqn:sm}), along the line of sight toward a source at
$(\ell,b)$ to form \hbox{SM}.  Since the two components have unequal
fluctuation parameters, the contribution from each component is
determined separately, then summed to form the total scattering
measure.  The predicted angular diameter is given by
equation~(\ref{eqn:scattangle}) where the modelled SM is a function
with the following parameters $\widehat{\mathrm{SM}} =
\widehat{\mathrm{SM}}(n_1, f_4, A_1, \hin, \hout, \fin, \fout,
\ell_0^\prime, \Psi| \ell, b; R_{\mathrm{warp}})$.

\section{Analysis}\label{sec:analyze}

In this section we describe how we have used the measured angular
diameters (\cite{lc97}) to obtain scattering diameters, discuss additional
scattering measurements we have used in our analysis, develop the
likelihood functions we will use to constrain the model parameters,
and present the results of this likelihood analysis.

\subsection{Determination of Scattering Diameters}\label{sec:sizes}

For each source we have a measurement of its apparent angular diameter
at one to three frequencies, $\thapp(\nu)$.  We assume the apparent
diameter is a quadrature sum of the intrinsic and scattering diameters
and model it as
\begin{equation}
\thapp^2(\nu) 
 = \frac{\thi^2}{\nu^{2\alpha}} 
 + \frac{\ths^2}{\nu^{4.4}}.
\label{eqn:appsize}
\end{equation}
Here $\thi$ and $\ths$ are the intrinsic and scattering diameters at
1~GHz, respectively.  For the scattering diameter we have used the
$\nu^{-2.2}$ dependence, as is appropriate for moderately strong
scattering.  For a homogeneous source with a peak brightness
temperature $T_B$, $\alpha = 1$ (\cite{ko88}).

We have considered a number of ways of using
equation~(\ref{eqn:appsize}) to solve for $\ths$.
\begin{enumerate}
\item Ignore intrinsic size: Setting $\thi = 0$~mas, $\ths =
\thapp\,\nu_{\mathrm{GHz}}^{2.2}$.  Since this method assumes that
scattering dominates the apparent diameter, the scattering diameter so
derived is an upper limit.

\item Dual frequency measurements: For sources with angular diameters
measured at 18 and~90~cm, if we set $\alpha = 1$, we have two
equations in two unknowns and can solve for $\thi$ and $\ths$.

\item Ignore scattering at high frequencies and assume homogeneous sources:
At 6~cm, scattering should be unimportant for lines of sight to the
outer Galaxy.  For those sources detected at 6~cm, we take the 6~cm
diameter to be the intrinsic diameter and scale it to 1~GHz, with
$\alpha = 1$.  We use the \emph{observed} frequency dependence to find
the 1~GHz apparent diameter from the 18 and 90~cm diameters.  The
scattering diameter is then found by subtracting in quadrature the
scaled intrinsic diameter from the apparent diameter.

\item Ignore scattering at high frequencies: The final method also
utilizes 6~cm diameters.  Rather than assuming $\alpha = 1$ in scaling
the intrinsic diameter to 1~GHz, we solve for $\alpha$ using the 6
and~18~cm diameters, assuming $\ths = 0$~mas.  Then, using the 18
and~90~cm diameters to solve for the 1~GHz apparent diameter, we again
subtract in quadrature the scaled intrinsic diameter from the apparent
diameter.
\end{enumerate}

Clearly not all of these methods can be used for all sources.  For
those sources for which multiple methods can be used, we utilize as
many of the methods as possible and then adopt the estimate which
places the most stringent limit on $\ths$.  In general, method~2 and
method~1 produce estimates of $\ths$ that are the same within the
errors.  We note that using method~1 tends to bias us toward larger
disk scale lengths because this method assumes that the intrinsic size
makes no contribution to the measured diameter.  We find 1~GHz
scattering diameters or limits in the range 1.5--48~mas; these are
tabulated in Table~\ref{tab:broaden}.

\subsection{Available Data}\label{sec:data}

We augment the scattering measurements from our survey (\cite{lc97})
with others from the literature within the same longitude range,
$150\arcdeg \le \ell \le 210\arcdeg$.  These are summarized in
Table~\ref{tab:literature}, scaled to 1~GHz.  Two kinds of scattering
measurements were found in the literature: angular broadening
measurements similar to those reported here and scintillation
bandwidth, $\Delta\nu_{\mathrm{d}}$, measurements of pulsars
(\cite{c86}).

The resulting data set consists of three classes of sources having the
following observables:
\begin{enumerate}
\item Extragalactic sources having measured scattering diameters or upper
limits on the scattering diameter, $\ths$; 11 such sources;

\item Pulsars with DM-independent distance estimates.  This class has
only one member, the Crab pulsar, for which a DM and scattering diameter
have been measured.

\item Pulsars without DM-independent distance estimates; 10 such
sources with measured DM and $\delnud$.
\end{enumerate}
All of these sources are shown in Fig.~\ref{fig:vlbalocs}.

Two of the pulsars that have a measured scintillation bandwidth will
not be included in our analysis.  The lines of sight to the pulsars
PSR~B0823$+$26 and PSR~B1112$+$50 are likely to be dominated by local
scattering material.  Both pulsars are closer than 0.5~kpc, closer
than the inner edge of the Perseus arm.

Our intention is to constrain the distribution of scattering material
in the outer Galaxy.  In selecting extragalactic sources to include in
our analysis, we have focussed on sources for which scattering makes a
measurable contribution the observed diameter.  Such sources are
marked by visibility functions displaying a gaussian-like profile with
increasing baseline length or by images containing only simple
structures, typically a single gaussian component.  An alternate
approach would be include in our analysis \emph{all} extragalactic
sources in the anticenter.  Since the observed angular size is a
convolution of the intrinsic size with the scattering diameter, we can
always derive an upper limit to the scattering diameter for any source
(see Method~1 in \S\ref{sec:sizes}).  However, these upper limits are
usually factors of at least 5--10 larger than the scattering diameters
predicted by the \cite{tc93} model.  Sources for which scattering appears to
dominate the observed angular diameters suggest that the level of
scattering toward the anticenter is actually less than that predicted
by the \cite{tc93} model.  The upper limits for the scattering diameters of
most sources therefore place no meaningful constraint on the
scattering toward the anticenter (see also \S\ref{sec:bestmodel}).  We
illustrate the lack of constraints provided by most sources with four
sources originally included in the \cite{tc93} analysis, but not included in
this analysis.

Four extragalactic sources included in the \cite{tc93} analysis are not
included here, because a re-analysis of the existing observations
suggests that no scattering diameter has been measured.  The four
sources are CTA21 (0316$+$162), 0611$+$131, 4C14.18 (0622$+$147), and
3C190 (0758$+$143).  In the \cite{tc93} analysis CTA21 and 3C190 were taken
to have scattering diameters of approximately 200~mas at~74~MHz, based
on a single-baseline VLBI experiment (\cite{r74}).  A later
multi-station VLBI experiment at 609~MHz showed CTA21 to have a
core-halo structure, with the halo having a size $\theta \ge 130$~mas
(\cite{wrap79}).  Wilkinson et al.~(1979) suggest that the halo is
responsible for the interplanetary scintillation this source exhibits
at~81~MHz.  The halo is also likely to be the component responsible
for the aforementioned 74~MHz angular diameter.  The core itself
appears to be a blend of two components.  A characteristic size of
these blended components is about 10~mas, equivalent to an upper limit
on the scattering diameter of 3.4~mas at~1~GHz; this upper limit is
more than a factor of~5 larger than the predicted \cite{tc93} scattering
diameter.  For 3C190, a later multi-station VLBI experiment at~609~MHz
showed it to have three components of comparable flux and similar,
steep spectra with diameters of approximately 100~mas (\cite{rsff91}).
It is not clear if one of these components dominates at~74~MHz, and
hence is responsible for the aforementioned angular diameter
measurement, or if Resch's~(1974) observations sample a complex
visibility function and his results do not represent an actual
diameter at all.  The equivalent upper limit on the 1~GHz scattering
diameter is approximately 40~mas, nearly a factor of 100 larger than
the predicted model diameter.  The source~4C14.18 has a steep spectrum
and a non-gaussian visibility function (\cite{dtbbbc84}).  By fitting
a gaussian to the visibility data, they place an upper limit on the
1~GHz scattering diameter of 22~mas, more than a factor of~5 larger
than that predicted by the \cite{tc93} model.  The source~0611$+$131 was
taken to have a scattering diameter of 40~mas at~408~MHz
(\cite{dtbbbc84}) in the \cite{tc93} analysis.  Our new observations at~1.6
and~5~GHz show this diameter to result from the blending of at least
two source components (\cite{lc97}).  We place an upper limit on the
1~GHz scattering diameter of 30~mas, a factor of~10 larger than the
predicted value.

Our initial attempts to form the likelihood function included the Crab
pulsar (PSR~B0531$+$21).  However, we found that it ended up
dominating the resulting likelihood functions, in some cases
contributing as much as 50\% of the log-likelihood.  Because the Crab
is relatively nearby ($D \approx 2$~kpc), we do not believe it should
be the dominant source in the likelihood function.  The results
presented here do not include the Crab.

The reason for the Crab's large contribution to the likelihood is a
large discrepancy between the observed and modelled quantities: The
observable quantities are $\mathrm{DM} = 56.8$~pc~cm${}^{-3}$ and
$\ths = 0.5 \pm 0.05$~mas, while typical values for these quantities
in our models are $\widehat{\mathrm{DM}} \approx 30$~pc~cm${}^{-3}$
and $\hat{\ths} \approx 0.7$~mas.  Our modelled values are comparable
to those in the \cite{tc93} model.  The models overpredict the
scattering angle, while underpredicting the \hbox{DM}.  In particular,
the discrepancy in the observed \textit{vs.}\ modelled DM results in a
significant contribution to the likelihood function.

The Crab nebula is unlikely to be the source of the discrepancies.
Its contribution to the DM is probably no more than 1\% (\cite{i77}).
Its contribution to the scattering diameter is de-leveraged by a
factor of order $L/D \sim 10^{-3}$ where $L$ is the diameter of the
nebula.  Evidence supporting a small nebular contribution to the
scattering diameter comes from comparing the measured scattering
diameter with that inferred from pulse broadening for the pulsar.  The
pulse broadening has a variable and a constant component; the constant
contribution arises from the ISM distributed between the pulsar and
the Earth.  The scattering diameter estimated from the constant pulse
broadening agrees well with the observed scattering diameter
(\cite{v76}; \cite{ir77}; Gwinn, Bartel, \& Cordes~1993).  We conclude
that the ionized gas along the line of sight to the Crab has a
relatively high electron density but is not strongly turbulent.

We are left with a total of 18 sources, 11 extragalactic and 7
pulsars.

\subsection{Likelihood Functions for Scattering in the Outer Galaxy}
\label{sec:likeli}

We shall use a likelihood analysis to constrain the parameters of the
ionized disk model presented in \S\ref{sec:ac.model}.

For the $i^{\mathrm{th}}$ line of sight, the probability of obtaining
the observable $x$ ($\ths$, DM, or $\delnud$) is 
\begin{eqnarray}
p\left(x_i | \hat{x}_i\right) & \approx & f_x\left(x_i | \hat{x}_i\right)\delta x_i \nonumber \\ 
 & = &\frac{1}{\sqrt{2\pi}}\exp\left[-\frac{1}{2}\left(\frac{x_i-\hat{x}_i}{\delta x_i}\right)^2\right].
\label{eqn:mprop}
\end{eqnarray}
where $\hat{x}_i$ is the value predicted for that line of sight and
$\delta x_i$ is the uncertainty associated with the measured value of
the observable.  For many of the extragalactic sources we have only an
upper limit to a scattering diameter, because the scattering diameter
has been estimated from a single frequency.  The probability that a
scattering observable $x$ is less than an upper limit $X$ is
\begin{eqnarray}
p\left(x_i \le X_i | \hat{x}_i\right)
 & = & \int_0^{X_i} dx^\prime\,f_x\left(x^\prime | \hat{x}_i\right) \nonumber \\
 & = & \frac{1}{2}\left[\erf\left(\frac{X_i-\hat{x}_i}{\delta x_i\sqrt{2}}\right)+ \erf\left(\frac{\hat{x}_i}{\delta x_i\sqrt{2}}\right)\right]
\label{eqn:ulprob}
\end{eqnarray}
where $\erf(x)$ is the error function.

The global likelihood function for all sources is
\begin{equation}
P = \prod_{i=1}^N p_i.
\label{eqn:global}
\end{equation}

The modelled quantities and measurement uncertainties for the
various classes of sources are as follow:
\begin{enumerate}
\item The predicted extragalactic scattering diameters are found by
integrating $d\mathrm{SM}$ along the line of sight and using
equation~(\ref{eqn:sm}).  Most of the extragalactic source scattering
diameters in the literature are single frequency measurements; we
scale the errors for these to 1~GHz assuming a $\lambda^{2.2}$
dependence.  For the scattering diameters we report, the errors are
estimated from the formal statistical errors on the fits to the data
and then scaled (single-frequency measurement) or propagated
(multiple-frequency determination) to 1~GHz.  The uncertainties are
10--25\%.

\item For the pulsars we integrate $n_{\mathrm{e}}\,ds$ and a
weighted form of equation~(\ref{eqn:sm}) until the modelled DM equals
the measured \hbox{DM}.  The appropriate weighting factor for the
scintillation bandwidth is $w(s) = (s/D)(D - s)/D$, where $D$ is the
distance to the source and $s$ is the distance along the line of
sight.  This weighting factor accounts for the fact that $\delnud$ is
a measure of the excess time delay taken by scattered lines of
sight.  The scintillation bandwidth is then calculated as 
\begin{equation}
\delnud 
 = 145\,\mathrm{Hz}\,D_{\mathrm{kpc}}^{-1}(\mathrm{SM})^{-6/5},
\label{eqn:ac.delnud}
\end{equation}
with SM in the conventional units of \smu\ and $D_{\mathrm{kpc}}$ the
distance in kpc.  The measurement uncertainties for $\delnud$ range
from 20\% to 70\%; we adopt $\delta(\delnud)/\delnud = 0.33$.
\end{enumerate}

\subsection{Results}\label{sec:results}

We have searched for the maximum of the global likelihood,
equation~(\ref{eqn:global}), using an iterative grid search procedure.
Initial ranges for the various parameters, described in more detail
below, were estimated based on the \cite{tc93} model results and the
structure of the \ion{H}{1} disk.  Then the ranges and the grid search
resolutions were refined to locate the maximum.

The model in \S\ref{sec:ac.model} contains nine parameters: $f_4$,
$n_1$, $A_1$, $\hin$, $\hout$, $\fin$, $\fout$, $\ell_0^\prime$, and
$\Psi$.  We held $n_1$ and $\hin$ fixed at their values in the
\cite{tc93} model, $n_1\hin = 0.0165$~kpc~cm${}^{-3}$ and $\hin =
0.88$~kpc ($n_1 = 0.0188$~cm${}^{-3}$).  These quantities are
constrained by the DMs of high-latitude pulsars, particularly those in
globular clusters.  There is only one high-latitude pulsar, B0301$+$19
($b = -33\arcdeg$) in our sample, and it is not in a globular cluster.
Further, $n_1$ and $\hin$ describe the ionized medium near the solar
circle and were constrained (along with nine other parameters) by a
fit to nearly 300 DM and SM measurements.  As we discuss at the end of
\S\ref{sec:ac.model}, we do not expect our more limited set of
measurements to change appreciably those parameters which describe the
electron density distribution in the inner Galaxy or even near the
solar circle.

We show that $\fin$ may have a value different than that adopted in
the \cite{tc93} model.  Because $\fin$ also affects sources in the solar
neighborhood that we do not include in our data sample, we shall not
conduct a grid search over it, but illustrate its effects by adopting
one of two fiducial values.

The existence and amplitude of the warp are modelled by $\Psi$ and
$\ell_0^\prime$.  From Fig.~\ref{fig:vlbalocs} it is apparent
that the current complement of anticenter scattering measurements
sample the outer Galaxy both coarsely and far from uniformly.  In
particular the sources are in three groups, with $\ell \approx
160\arcdeg$, 180\arcdeg--190\arcdeg, and 200\arcdeg.  The number
of sources ($\approx 20$) available for constraining the shape of
the outer \ion{H}{2} disk is also far smaller than the number used
to describe the \ion{H}{1} warp ($\sim 40\,000$ telescope beams,
\cite{btlh86}), the $\mathrm{H}_2$ warp ($\sim 1300$ IRAS sources,
\cite{wbbk90}), or the stellar warp ($\sim 20\,000$--90\,000
stars, \cite{ds89}; \cite{bwcps93}).  Our survey is more
restricted in longitude than these other surveys, but, even so,
the above surveys have 10--100 times as many lines of sight.  We
shall therefore not conduct a search over $\Psi$ or
$\ell_0^\prime$ but shall choose two fiducial pairs and compare
the resulting maximum likelihood values.  One pair will be for an
unwarped disk, $\Psi = 0\arcdeg$ ($\ell_0^\prime$ is, of course,
undefined for an unwarped disk), and the second pair will be
characteristic of the warp in the \ion{H}{1} disk, $(\Psi,
\ell_0^\prime) = (20\arcdeg, 170\arcdeg)$.

In summary, we never attempted to fit for more than three parameters,
$A_1$, $\hout$, and $\fout$ at a time.
In the rest of this section, we place preliminary constraints on
$A_1$, assess the importance of the Perseus arm, then reevaluate our
constraints on $A_1$.

\subsubsection{Preliminary Constraints on $A_1$}

Figure~\ref{fig:disk} shows the likelihood as a function of $A_1$ for
a model with an unwarped, non-flaring disk and in which the Perseus
arm does not contribute to the scattering, i.e., $f_4 = 0$.  

We have excluded PSR~B0611$+$22 and the extragalactic source
0629$+$109 from this fit.  The pulsar we exclude because its DM and
scintillation bandwidth are likely to be affected by the Str{\"o}mgren
spheres of stars in the I~Gem association (Weisberg, Rankin, \&
Boriakoff~1980).  The extragalactic source we exclude because it has a
1~GHz diameter of 25~mas (\cite{dtbbbc84}), approximately a factor of
five larger than that for extragalactic sources within a few degrees.
The line of sight for this source passes close to the edge of the
\ion{H}{2} region Sharpless~273, $(\ell,b) \approx
(202\arcdeg,2\arcdeg)$.  This \ion{H}{2} region probably enhances the
scattering for this line of sight.  We retain these sources in the
fits which include the spiral arm component (below) because the spiral
arms have been incorporated in the \cite{tc93} model specifically to account
for enhanced dispersion and scattering such as would occur from OB
associations and \ion{H}{2} regions.

The maximum likelihood occurs at $A_1 \approx 17$~kpc for a $\sech^2$
disk while $A_1 \approx 25$~kpc for a truncated disk.  The truncated
disk is larger because there is no scattering material outside $A_1$
while, for the $\sech^2$ dependence, $A_1$ is approximately the
half-power point and there is a non-negligible amount of ionized gas
at $2A_1$ ($\approx 10$\%).  The likelihood favors the $\sech^2$ disk,
but by a factor less than two.

Provided that $\fout$ is not substantially smaller than $\fin$, these
likelihood results place an upper limit on $A_1$.  Any scattering
contributed by the Perseus arm would reduce the estimate of $A_1$ (as
the second panel of Fig.~\ref{fig:disk} illustrates and we discuss
below).  Similarly, allowing the disk to flare or warp or both
produces, on average, larger scattering diameters for high-latitude
sources.  To reproduce a given scattering diameter in the absence of
flaring or warping, the scattering material must extend to a large
enough distance, i.e., $A_1$ must be large enough, to compensate for
the $z$-dependent fall-off of the scattering material.  As we
demonstrate below, however, if $\fout$ is quite small, as compared to
$\fin$, a much larger radial extent would be favored.

On the basis of Fig.~\ref{fig:disk}, we conclude that the large
$A_1$ values that were allowed by the \cite{tc93} analysis, e.g., $A_1
\approx 50$~kpc, are now shown to be disfavored.

\subsubsection{Scattering in the Perseus Arm}

The nominal Perseus arm in the \cite{tc93} model contributes enough
scattering that some scattering diameters are overpredicted by a
factor of two.  Thus, we must decrease the amount of scattering
contributed by this arm.

In the \cite{tc93} model this arm has a fluctuation parameter equal to
that of all other arms, $F_{\mathrm{a}} = 6$, and a fine-tuning
parameter $f_4 = 1$.  Our model allows $f_4$ to vary.  Since the
Perseus arm scattering measure is $\mathrm{SM} \propto
F_{\mathrm{a}}f_4^2$, adjusting $f_4$ is equivalent to allowing the
various arms to have different fluctuation parameters.

If we set $n_1 = 0$~cm${}^{-3}$, thereby ``turning off'' the scattering
in the disk, we determine how large $f_4$ must be to account for all
of the scattering required by the observed scattering diameters.  We have
performed two separate fits for $f_4$, one using all of the sources,
the second using just the extragalactic sources.  Since the
extragalactic sources are clearly well beyond all of the Galaxy's
scattering material, the extragalactic sources should provide an upper
limit to $f_4$.  In both cases we find $f_4 \approx 0.6$.  Allowing
the disk to contribute to the scattering will require an even smaller
$f_4$.  

The estimate for $A_1$ in the previous section assumed $f_4 = 0$.
Figure~\ref{fig:disk} also shows the estimate of $A_1$ obtained for
$f_4 = 0.25$, with PSR~B0611$+$22 and 0629$+$109 included in the
fitting.  Because of the additional scattering contributed by the arm,
the estimates of $A_1$ are smaller, $A_1 \lesssim 15$~kpc for a
$\sech^2$ dependence and $A_1 \lesssim 22$~kpc for a truncated disk.
The likelihood ratio between the $\sech^2$ and truncated disk again
favors the $\sech^2$ disk by a factor of less than two.

In the following we reconsider our limit on $A_1$ while allowing the
disk to flare and warp.  Rather than conduct a search over the range
$0 \le f_4 \lesssim 0.6$, we shall evaluate the likelihood function at
a fiducial value.  Comparing the magnitudes of the likelihood
functions for $f_4 = 0$ and $f_4 = 0.6$, the value $f_4 = 0$ is
favored by more than an order of magnitude.  We adopt $f_4 = 0.25$.
With this value of $f_4$, the Perseus arm contributes an $\mathrm{SM}
\sim 6.25 \times 10^{-4}$~\smu, equivalent to a 1~GHz scattering
diameter of 1.5~mas.  Figure~\ref{fig:disk} indicates how a different
choice for $f_4$ would affect the maximum likelihood value of $A_1$.

\subsubsection{The Outer Ionized Disk}\label{sec:ac.results}

In addition to $A_1$ we shall be fitting for $\hout$ and $\fout$.  Our
initial ranges were centered approximately on the values in the
\cite{tc93} model of 0.88~kpc and 0.4, respectively.  Our initial
range for $\hout$ was $0.5\,\mathrm{kpc} \le \hout \le
5\,\mathrm{kpc}$.  The upper limit is comparable to the scale height
for \ion{H}{1} (\cite{btlh86}); the lower limit is approximately half
the inner scale height, $\hin = 0.88$~kpc.  Further, the
\cite{tc93}-model value for the scale height of the material in the
spiral arms is $h_{\mathrm{a}} = 0.3$~kpc.  Values of $\hin \lesssim
0.5$~kpc could indicate an underestimate of the scattering in the
Perseus arm, i.e., $f_4$ is too low.  Our initial range for $\fout$
was $0 \le \fout \le 2$.  The lower limit describes a quiescent, i.e.,
non-scattering, outer disk.  The fluctuation parameter for the spiral
arms is $F_{\mathrm{a}} = 6$ so fitted values of $\fout \gtrsim 2$
could also indicate that the scattering contribution from the Perseus
arm was underestimated.

Figure~\ref{fig:like7} compares the likelihoods for a $\sech^2$ radial
dependence \textit{vs.}\ a truncated disk for an unwarped disk.  The primary
difference between these two models is that $A_1 \approx 22$~kpc for a
truncated disk \textit{vs.}\ $A_1 \approx 15$~kpc for the $\sech^2$ disk.
These differences are comparable to those found	in the more simple model of
Fig.~\ref{fig:disk}.

The likelihood results favor $\hout \approx 0.6$~kpc, as compared to
value of 0.88~kpc for the inner Galaxy from the \cite{tc93} model.
Above we identify an underestimate of the level of scattering in the
Perseus arm as one means of producing $\hout \lesssim \hin$.  We now
identify three additional possibilities: (1)~Misclassification of a
source as extragalactic rather than Galactic; (2)~Less scattering near
the solar circle than predicted by the nominal \cite{tc93} model; or
(3)~A patchy distribution of scattering material.

If a source is classified as extragalactic, the model value of SM is
calculated from an integral along the entire line of sight.  If the
source were instead classified as Galactic, the integral would extend
out only to the source's estimated distance.  Not only would a shorter
path length result in a smaller SM and smaller predicted diameter, but the
path would sample less of the material at large $z$.  The scale height
would be accordingly less constrained.

The diameter of the source 87GB~0600$+$2957 ($b = 4\fdg0$) is 1.5~mas.
The predicted diameter of the source in the unwarped disk models
described above is approximately 2.5~mas.  The predicted diameter from
the \cite{tc93} model is nearly 6~mas.  If we repeat the above
fitting, excluding 87GB~0600$+$2957 from the sample of sources, the
maximum likelihood estimate of $\hout$ nearly doubles, while the
estimates of $A_1$ and $\fout$ remain essentially unchanged.  However,
as we discuss in \cite{lc97}, we can find no compelling reason to
classify this source as Galactic.

The second possibility is that the \cite{tc93} model may overestimate
the level of scattering toward the anticenter.  \cite{tc93} estimated
the fluctuation parameter in the extended component to be $\fin =
0.36^{+0.30}_{-0.10}$ and adopted a nominal value of $\fin = 0.4$.  We
have repeated the analysis above with $\fin = 0.3$.  The likelihood
results are essentially unchanged, as we might expect.  This lower
value of $\fin$ is a reduction of only 25\%, while the discrepancy
between the observed and modelled diameters for 87GB~0600$+$2957 is
nearly a factor of two.  A lower value of $\fin$ can reduce, but not
eliminate, this discrepancy.  We therefore conclude that the
anticenter distribution of scattering material may contain holes or
gaps---a source shining through one of these gaps would have an
anomalously small scattering diameter.

Harrison \& Lyne~(1993) compared the velocities as determined by
proper motions and interstellar scintillation pattern velocities for a
number of high latitude pulsars.  They concluded that the scale height
of the scattering gas and of the ionized gas were markedly different:
approximately 0.1~kpc for the former and 1~kpc for the latter.  We
assume that the scattering traces the distribution of ionized gas,
however, our analysis considers only the scattering data.  Regardless
of the correct explanation for the small angular diameter of
87GB~0600$+$2957, we strongly disfavor a scale height as small as
0.1~kpc and find the scale height to be 5--10 times larger.

Figure~\ref{fig:likee}\textit{a} shows the likelihood function for a
warped disk, constructed using the observables from all 18
sources. The estimates of $A_1$ change only slightly from the unwarped
disk.  However, the estimates of $\hout$ increase by more than a
factor of five.  Larger values of $\hout$ are allowed primarily
because of our coarse angular sampling.  Scattering observables for
sources near $\ell \approx 180\arcdeg$ are little affected by a warp
in the disk.  The warp is such that the maximum electron column
density shifts to positive latitudes in the second quadrant and
negative latitudes in the third.  Examination of
Fig.~\ref{fig:vlbalocs} shows that the warp generally increases the
angular distance between sources and the midplane of the disk.  Thus,
larger scale heights are allowed, indeed required, in order to
reproduce the observed source diameters.

In Fig.~\ref{fig:likee}\textit{b}, we assess the influence of the
pulsars on the likelihood functions.  The pulsars' contribution to the
likelihood is constructed from the observed and modelled scintillation
bandwidth, $\delnud$ and $\widehat{\delnud}$, respectively.  In turn,
$\widehat{\delnud}$ is a function of both $\widehat{\mathrm{SM}}$ and
$\widehat{\mathrm{DM}}$, equation~(\ref{eqn:ac.delnud}).  The
$\widehat{\mathrm{DM}}$ dependence occurs because $\widehat{\delnud}$
and the weighting function for $\widehat{\mathrm{SM}}$ both depend
upon distance, which is estimated by integrating $n_{\mathrm{e}}\,ds$
until it equals the observed \hbox{DM}.  In contrast, the scattering
diameter for an extragalactic source is a function of only
$\widehat{\mathrm{SM}}$, equation~(\ref{eqn:scattangle}).  Hence
systematic errors in the model have a greater impact on the pulsars'
contribution to the likelihood.

The regions of maximum likelihood in Figs.~\ref{fig:likee}\textit{a}
and~\ref{fig:likee}\textit{b} overlap, though the allowed regions in
Fig.~\ref{fig:likee}\textit{b} are somewhat larger.  The larger
regions are to be expected since fewer sources were used to constrain
the model parameters.  Comparing the left and right panels of
Fig.~\ref{fig:likee}, the most significant difference is in the
maximum likelihood estimates for $A_1$ and $\fout$.  The m.l.e.\ for
$A_1$ is smaller and the m.l.e.\ for $\fout$ is larger when only
extragalactic sources are used as compared to all sources.  This
difference is due primarily to the presence of 0629$+$109 in the
sample.  Its scattering diameter is large enough that considerable
scattering, i.e., large $\fout$, is needed.  However, the smaller
diameters of the other extragalactic sources then drive $A_1$ to
smaller values in order that their scattering diameters not be
overestimated.  We have also repeated this analysis excluding
0629$+$109 from the sample.  In this case the likelihood functions
including and excluding the pulsars are nearly indistinguishable,
indicating that the contribution of the extragalactic sources to the
likelihood function dominates that of the pulsars.

\subsection{Preferred Model}\label{sec:bestmodel}

Our likelihood results favor the unwarped disk over the warped disk by
a factor of 5--10.  In both models, the $\sech^2$ and truncated radial
dependences are nearly equally likely.  We favor the truncated disk
because its radial extent is in good agreement with the radial extent
of sites of massive star formation,
\S\ref{sec:ac.ionize} (cf.\ Fig.~\ref{fig:ac.summary}).

The nominal set of parameters can be obtained largely by inspection of
Fig.~\ref{fig:like7} and is summarized in Table~\ref{tab:params}.  We
adopt $A_1 = 20$~kpc and $\fout = 0.4$, the latter is slightly less
than what Fig.~\ref{fig:like7} suggests, but is in good agreement with
the value that \cite{tc93} derive for the inner Galaxy.  Thus, the
extended scattering component has a continuous fluctuation parameter
through the onset of the warp.  For the scale height $\hout$ we adopt
1~kpc.  This is larger than what Fig.~\ref{fig:like7} suggests, but,
as we discuss above, the estimate of $\hout$ is influenced by
87GB~0600$+$2957.  A value of 1~kpc is intermediate between that value
derived using 87GB~0600$+$2957 in the fitting and that value derived with
87GB~0600$+$2957 omitted from the fitting.

In Fig.~\ref{fig:decon} we show the contribution of the individual
sources to the total likelihood of Fig.~\ref{fig:like7} (cf. also
Fig.~\ref{fig:disk}).  This figure can be used to assess how well our
model reproduces the observed scattering diameter.  The fact that we
have been able to limit $A_1$ has been due largely to the addition of
two sources, 87GB~0600$+$2957 and 87GB~0621$+$1219 (\cite{lc97}).  The
remaining sources place either a lower limit on $A_1$ or do not
constrain it very well at all.  One source not shown on this figure is
0629$+$109.  Its contribution to the likelihood is significantly less
than the other sources.  Since the line of sight to 0629$+$109 passes
close to the \ion{H}{2} region S273, this low likelihood could
indicate either that our adopted value for $f_4$ is too small or that
the assumption that spiral arms are smooth structures is beginning to
break down.

This figure also demonstrates why using upper limits to the scattering
diameters of all extragalactic sources in the anticenter would not be
worthwhile, \S\ref{sec:data}.  Sources with upper limits to the
scattering diameter significantly larger than the modelled value would
produce on this plot flat lines coincident with the $\log p = 0$ abscissa.

\section{Discussion and Conclusions}\label{sec:conclude}

\subsection{Free Electrons and Turbulence in the Outer Galaxy}

Our likelihood results indicate that $A_1 \approx 15$~kpc if the
fluctuating part of the electron density distribution exhibits a
gradual decrease in the outer Galaxy or $A_1 \approx 20$~kpc if the
\ion{H}{2} disk is truncated.  The apparent decrease in the rms
$n_{\mathrm{e}}$ could result if sites of turbulence became less
numerous while the mean $n_{\mathrm{e}}$ remained constant or it could
result from decreases in the mean $n_{\mathrm{e}}$ itself.  In this
section we assess the extent to which our inferred value for $A_1$ can
distinguish between these possibilities.

To discuss variations in the electron density, we recast
equation~(\ref{eqn:sm}) in terms of the \ion{H}{1} density and
ionization and integrate over a path length $D$,
\begin{equation}
\mathrm{SM}
 = \int_0^D C_{\mathrm{u}}FfX_{\mathrm{i}}^2n_{\mathrm{H}}^2\,ds.
\label{eqn:sm_ii}
\end{equation}
In this form, it is clear that there are three means by which the amount
of scattering in the outer Galaxy could be limited:
\begin{enumerate}
\item $n_{\mathrm{H}}$---It is well known that the distribution of
\ion{H}{1} decreases at large Galactocentric radii (\cite{b92}).
Outer Galaxy scattering could be limited because there is simply not
enough gas to be ionized and produce scattering.

\item $fX_{\mathrm{i}}^2$---This factor is the product of the fractional
ionization, $X_{\mathrm{i}}$, and the volume filling factor, $f$, of
the ionized gas.  If the number density of ionization sources, e.g.,
\ion{H}{2} regions and supernovae, decreases faster than does
$n_{\mathrm{H}}$, the scattering would be limited by a lack of ionized
gas, even though there would be sufficient amounts of neutral gas.  We
treat the product $fX_{\mathrm{i}}^2$ rather than $f$ and
$X_{\mathrm{i}}$ separately because our measurements cannot
distinguish between the two. Changes in $f$ can be balanced by changes
in $X_{\mathrm{i}}^2$ so as to keep the product constant.

\item $F$---This factor is a measure of the level of turbulence.
Since the rate of star formation does fall off toward the outer
Galaxy, unless alternate or additional sources of ionization are
present in the outer Galaxy, e.g., the intergalactic ionizing flux,
the scattering could be limited by a lack of turbulent energy input
into the medium rather than the ionization.
\end{enumerate}
Our motivation for assuming that SM is separable in this manner is to
consider ionization and energy input mechanisms from sources not
generally thought to be operative in the inner Galaxy.  We consider
each of these factors in turn.

\subsubsection{Hydrogen Distribution}

Outside the solar circle atomic hydrogen dominates molecular hydrogen
(\cite{gb76}) and we take $n_{\mathrm{H}} = n_{\mion{H}{1}}$.
Comparison of the emission measure ($\mathrm{EM} = \int
n_{\mathrm{e}}^2\,ds$) and DM toward high-latitude pulsars suggest $f
\gtrsim 0.1$ and $X_{\mathrm{i}} \approx 1$ (\cite{r77}).  Because we
are assuming the strength of scattering in the outer Galaxy is
dominated by the decrease in the \ion{H}{1} density, we take $f$,
$X_{\mathrm{i}}$, and $F$ to be constant as a function of $R$.

We estimate the quantity $\int n_{\mion{H}{1}}^2\,ds$ toward the
anticenter using the mass models of Dehnen \& Binney~(1997).  We are
required to use a model to estimate this quantity because the
self-opacity of \ion{H}{1} toward the anticenter means that the
\ion{H}{1} column density is not an observable (\cite{btlh86}).

Our estimate is $\int n_{\mion{H}{1}}^2\,ds \approx 2.3$~cm${}^{-6}$~kpc.
Assuming that $F = 0.4$, the resulting SM is $\log_{10}(\mathrm{SM}) =
-0.8$.  This SM produces a 1~GHz scattering diameter of 40~mas,
approximately 2--5 times larger than what we observe.  We conclude
that the distribution of electron density turbulence must decrease
more rapidly with Galactocentric distance than the distribution of
hydrogen.  This decrease may be due to an overall decrease in ionized
gas or to a diminution of turbulence in the ionized gas.

\subsubsection{Sources of Ionization: Stellar \textit{vs.}\ Intergalactic}\label{sec:ac.ionize}

Interior to $R \approx 25$~kpc, the \ion{H}{1} surface density is
greater than $10^{19}$~cm${}^{-2}$ (\cite{b92}; \cite{db97}) and the
disk is optically thick to the intergalactic ionizing flux.
Consequently, $fX_{\mathrm{i}}^2$ should increase with $R$ as the
\ion{H}{1} surface density and the disk's optical depth decrease.
Patchiness in the outer Galaxy \ion{H}{1} distribution would also
increase $fX_{\mathrm{i}}^2$.

As the previous section showed, if $fX_{\mathrm{i}}^2 \approx 0.1$ and
is constant as a function of $R$, the \ion{H}{1} distribution
overpredicts the scattering diameters.  Allowing $fX_{\mathrm{i}}^2$
to increase with $R$ would increase the size of this discrepancy.
Further, the scale length of the scattering is smaller than the Galactocentric
distance at which the disk becomes optically thin, indicating that the
intergalactic ionizing flux does not play a significant role in the
scattering in the outer Galaxy.

We favor star formation in the outer Galaxy as the more likely source
of ionization.  Wouterloot et al.~(1990) showed that the distribution
of molecular clouds with embedded massive star formation terminates at
20~kpc.  This truncated distribution has an extent comparable to what
our likelihood functions imply for the radial extent of scattering.
The limited radial extent of the molecular clouds is also the reason
we favor the truncated disk model to describe the radial dependence of
scattering.  Figure~\ref{fig:ac.summary} illustrates schematically the
spatially coincident distributions of molecular clouds and turbulent
gas in the outer Galaxy.

\subsubsection{Sources of Turbulence: Stellar \textit{vs.}\ Galactic}

The factoring of the scattering measure into separate ionization and
turbulent contributions, equation~(\ref{eqn:sm_ii}), ignores possible
correlations between these factors: Many of the same sources
responsible for the ionization of the gas can also serve to produce
turbulence, e.g., \ion{H}{2} regions and supernovae.  We utilize this
factoring in order to consider an alternate source of turbulence not
associated with massive star formation.  We conclude that turbulence
is in fact associated with star formation.

The orbits of the Magellanic clouds cause them to cross the midplane
of the disk.  As these and other satellite galaxies cross the disk,
they generate mixing layers and wakes.  The eddy turnover time, which
is related to the energy dissipation rate, is $t \sim l_0/u$
(\cite{tl72}) where $l_0$ is the outer scale of the turbulence and $u$
is a characteristic velocity.  For $l_0 \sim 100$~pc (\cite{r90};
\cite{s91}; and references within) and $u \sim 100$~km~s${}^{-1}$, $t
\sim 10^6$~yr.  Since the orbital period of a typical satellite galaxy
is of order $10^9$~yr, the passage of a satellite galaxy through the
outer disk will provide only a transitory source of turbulence.

The distribution of molecular clouds with embedded massive star
formation extends to 20~kpc (\cite{wbbk90}), comparable to the extent
of the extended ionized component.  The stars embedded in these
molecular clouds have spectral types of early B (Wouterloot, Brand, \&
Henkel~1988).  These stars are probably sufficiently powerful to
produce turbulence:  The pulsar B0611$+$22 exhibits
strong interstellar scintillation and the line of sight to it passes
near several late O and early B stars (\cite{wrb80}).

\subsection{Comparison with External Galaxies}

Diffuse ionized gas in external galaxies has been detected in
H$\alpha$ emission (\cite{r96} and references within).  This diffuse
gas is presumably the equivalent of the Galaxy's warm ionized medium
(\cite{kh87}).  Spangler \& Reynolds~(1990) showed that the scattering
diameters for extragalactic sources are correlated with H$\alpha$
emission, suggesting that the same (warm ionized) gas responsible for
the H$\alpha$ emission is also responsible for the scattering.  In
this section we compare the radial extent of the Galaxy as inferred in
our likelihood analysis with that determined for other galaxies.

Table~\ref{tab:external} presents a subset of H$\alpha$ measurements
extracted from the literature.  We restrict the list to those galaxies
for which the published observations include images large enough that
radial extents can be estimated reliably.  The last entry in
Table~\ref{tab:external} is our estimate, based on the likelihood
results, for the radial extent of the Galaxy as it would appear to an
external observer.  We have obtained this estimate in the following
manner.  The ($1\sigma$) sensitivities for the H$\alpha$ observations
are typically $\delta(\mathrm{EM}) \approx 3$~cm${}^{-6}$~pc.  We
integrated $n_{\mathrm{e}}^2\,ds$ to produce the modelled EM,
$\widehat{\mathrm{EM}}$, along a line of sight appropriate for an
external observer seeing the Galaxy edge on, i.e., along the path
parallel to the $x$-axis (the discussion following eqn.~[\ref{eqn:ne}]
describes the coordinate system).  Trial and error was sufficient to
determine that, for a truncated disk, $\widehat{\mathrm{EM}} \approx
\delta(\mathrm{EM})$ at $R \approx A_1/2$.  The radial extent of the
Galaxy is comparable to the radial extent of these other galaxies.

A measure of the star formation rate of the galaxy is provided by
$L_{\mathrm{FIR}}/D_{25}^2$, the far-infrared luminosity within
the optical isophotal diameter at $25^{\mathrm{th}}$ magnitude.
This is an imperfect measure of the star formation rate, however,
as low-mass stars heating ``cirrus'' clouds can contribute to the
far infrared luminosity (\cite{r96}) and the optical and infrared
luminosities may have different extents.

Although the star formation rate, as measured by
$L_{\mathrm{FIR}}/D_{25}^2$, is a good predictor of the amount of
extraplanar gas (\cite{r96}), it is not well correlated with the
radial extent of H$\alpha$.  For instance, the radial extent of
NGC~4217 is only 25\% (4~kpc) smaller than that of UGC~10288 even
though the star formation rate is at least a factor of three higher in
UGC~10288; NGC~5746 has a radial extent larger than that of the Galaxy
even though its star formation rate is an order of magnitude less than
the Galaxy's.  More likely, the star formation rate is determined by a
quantity like the \ion{H}{1} or H${}_2$ surface density.

Further support for our proposal that our measurements trace the
extent of the turbulent ionized disk is found by comparing the
H$\alpha$ emission in external galaxies (\cite{r96}, Figs.~1--9) with
the molecular cloud distribution in the outer Galaxy (\cite{wbbk90},
Fig.~4).  The H$\alpha$ emission is concentrated toward the galaxies'
centers with a gradient to large radial distances.  Near the edge of
the H$\alpha$ disk, the emission becomes patchy.  The molecular cloud
distribution in the outer Galaxy has a similar appearance---it display
a strong Galactocentric gradient and, for $R \approx 15$--20~kpc, the
distribution is patchy.

High latitude structure in the extended component could influence
our estimates for the various model parameters, in particular
$\hout$, \S\ref{sec:ac.results}.  The morphology of the
extraplanar diffuse gas in external galaxies shows considerable
variety: NGC~891 shows diffuse gas up to 4~kpc off the plane with
many vertical filaments (Rand, Kulkarni, \& Hester~1992);
UGC~10288 shows vertical filaments but no diffuse gas
(\cite{r96}); NGC~4278 shows patchy emission (\cite{r96}); and
NGC~4565 shows little halo diffuse gas (\cite{rkh92}).  H$\alpha$
observations of the diffuse gas in the solar neighborhood show
filamentary structure (\cite{or85}) and the vertical morphology of
the extended component could be quite complex.

\subsection{Conclusions}

We have modified the outer Galaxy portion of the Taylor-Cordes model
for the global distribution of ionized gas.  Our modifications are
motivated by the observed warping and flaring of the \ion{H}{1},
H${}_2$, and stellar constituents of the outer Galaxy.  The data
available to constrain the model consist of 18 sources, 9 
extragalactic sources from a survey we conducted (\cite{lc97}) and 7
pulsars and 2 extragalactic source extracted from the literature
(Table~\ref{tab:literature}).  We used a likelihood analysis to
constrain the model parameters.  The two most important parameters are
the radial scale length of the ionized disk, $A_1$ and the strength of
scattering in the Perseus arm.  The adopted model parameters are
summarized in Table~\ref{tab:params}.

The scattering in the Perseus arm is, at most, 60\% of the level seen
in the inner Galaxy spiral arms.  This upper limit assumes that all
of the scattering for sources toward $\ell \sim 180\arcdeg$ is due
to the Perseus arm.  Our analysis favors a level of scattering less
than this upper limit.  We adopt a value 25\% that in the inner
Galaxy; the equivalent scattering diameter is 1.5~mas at~1~GHz.

We considered two different radial dependences for the electron
density, a smooth decrease of the electron density with Galactocentric
distance, equation~(\ref{eqn:radial1}), and a truncated distribution,
equation~(\ref{eqn:radial2}).  The current data cannot distinguish
between these two forms.  The radial scale length for the ionized disk
is $A_1 \approx 15$--20~kpc, comparable to the extent that \cite{tc93}
adopted with fewer anticenter measurements.  We favor the truncated
disk because the radial extent inferred for sites of massive star
formation also appears truncated at approximately 20~kpc, as indicated
schematically in Fig.~\ref{fig:ac.summary}.  H$\alpha$ observations of
external galaxies show that they have radial extents comparable to
that which we infer for the Galaxy.

Our analysis favors an unwarped, non-flaring disk with a scale height
of 1~kpc, though this may reflect the non-uniform and coarse coverage
of the anticenter provided by the available data.

The observed scattering diameter of one extragalactic source
(87GB~0600$+$2957) is a factor of two smaller than the modelled
scattering diameter, suggesting the possibilities of holes in the
scattering material.  The H$\alpha$ emission in the outer portions of
the disks of these external galaxies also appears patchy, similar to
the distribution of molecular clouds in the outer Galaxy.  A patchy
distribution of massive star formation sites would allow the
possibility of holes in the scattering material.

We conclude that scattering in the outer Galaxy traces star formation,
as it does in the inner Galaxy.  The intergalactic ionizing flux and
turbulence generated by satellite galaxies passing through the disk
contribute little to the scattering.  However, the ionized disk of the
Galaxy could extend to much larger radii ($R \gtrsim 100$~kpc),
comparable to that inferred from Ly$\alpha$ absorption systems (e.g.,
\cite{csh93}), if the extreme outer disk is quiescent and contributes
little scattering.

\acknowledgements
We thank P.~Goldsmith, D.~Chernoff, and T.~Herter for helpful
conversations.  We thank the referee for a suggestion that led to
Fig.~\ref{fig:decon}.  This research has made use of the Simbad
database, operated at the CDS, Strasbourg, France.  This research was
supported by NASA GRO grants NAG~5-2436 and NAG~5-3515 and NSF grant
AST-9528394.

\clearpage

\begin{figure}
\caption[]{The angular broadening for an extragalactic source observed
toward the anticenter as a function of $A_1$, the Galactocentric scale
length in the Taylor-Cordes model.  The curves are labelled by the
observation frequency in GHz.  The nominal resolutions of the Very
Long Baseline Array and an array containing the HALCA VLBI satellite
are shown as dotted lines.}
\label{fig:angle}
\end{figure}

\begin{figure}
\caption[]{The functional forms for the radial dependence of
$n_{\mathrm{e}}$ in the outer Galaxy.  The solid line shows the radial
dependence of the \ion{H}{1} volume density (\cite{gb76}; \cite{b92});
the long-dash line shows the $\sech^2$ dependence of
equation~(\ref{eqn:radial1}); and the short-dash line shows the
truncated disk of equation~(\ref{eqn:radial2}).  All quantities are
normalized to pass through unity at $R = R_{\sun}$.}
\label{fig:radial}
\end{figure}

\begin{figure}
\caption[]{The angular distribution of low-latitude anticenter sources
with measured scattering observables.  The plotted symbol is
proportional to the logarithm of the 1~GHz scattering diameter.
Circles show scattering diameters from this program, squares show
scattering diameters for extragalactic sources reported in the
literature, and stars show the diameters of pulsars.  Except for the
Crab, $(\ell,b) = (184\arcdeg, -5\arcdeg)$, the pulsar diameters are
inferred from the pulsar's scintillation bandwidth and require an
estimate of the pulsar's distance.  The contour increments are 1~mas
with the highest contour being 7~mas.
\textit{Top}: The contours show the scattering diameters as predicted
by the Taylor-Cordes model. 
\textit{Bottom}: The contours show the scattering diameters as predicted by
a warped, non-flaring model for the disk.  The onset of the warp is at $R =
10.5$~kpc and the warped disk is inclined by $\Psi = 20\arcdeg$ to the
inner Galaxy's midplane, viz.\ Fig.~\ref{fig:ac.geometry}.}
\label{fig:vlbalocs}
\end{figure}

\begin{figure}
\caption[Tilted Ring Geometry for the Outer Ionized Disk]
{The geometry of the tilted ring model for the outer ionized disk.  A
source with Galactic coordinates $(\ell,b)$ has a latitude $b^\prime$
relative to the tilted ring's midplane.  The scale height of the disk
exterior to $R = 10.5$~kpc is also shown.}
\label{fig:ac.geometry}
\end{figure}

\begin{figure}
\caption[]{Likelihood estimates of $A_1$ for an unwarped, non-flaring
disk.  The solid line shows the likelihood results for a disk with a
radial $\sech^2$ dependence of equation~(\ref{eqn:radial1}), the
dashed line for a truncated disk of equation~(\ref{eqn:radial2}).  
\textit{Top}: The likelihood function if the Perseus spiral arm is
ignored.
\textit{Bottom}: The likelihood function if the Perseus spiral arm has a
strength 25\% that of the inner Galaxy spiral arms.}
\label{fig:disk}
\end{figure}

\begin{figure}
\caption[Outer Disk Likelihood Functions---Unwarped Disk]
{Likelihood function contours for pairs of parameters that include the
disk scale length, $A_1$, scale height, $\hout$, and fluctuation
parameter, $\fout$.  The disk is unwarped. Contours show the 67\%,
90\%, and 99\% confidence regions.  Crosses mark the location of
the maximum likelihood.
\textit{Left panels}:  $\sech^2$ radial dependence for the disk,
equation~(\ref{eqn:radial1}); and 
\textit{Right panels}: truncated disk,
equation~(\ref{eqn:radial2}).
The sharp edges in these likelihood functions are real and result from
fluctuations in the likelihood function caused by the small number of
data.  They do not reflect the grid resolution in the grid searches.}
\label{fig:like7}
\end{figure}

\begin{figure}
\caption[Outer Disk Likelihood Functions---Warped Disk]
{As for Fig.~\ref{fig:like7}.  The model disk is warped with $\Psi =
20\arcdeg$ and $\ell_0^\prime = 170\arcdeg$ and has a $\sech^2$
dependence.
\textit{Left}:  Likelihood function constructed using all data; and
\textit{Right}: Likelihood function constructed using only
extragalactic sources.}
\label{fig:likee}
\end{figure}

\begin{figure}
\caption[]{The contribution of the individual sources to the
likelihood function for the truncated disk model in
Fig.~\ref{fig:like7} (cf.\ also Fig.~\ref{fig:disk}).  We have held
all parameters fixed at their maximum likelihood estimated value
(viz.\ Table~\ref{tab:params}), except $A_1$.  Sources whose
individual likelihoods fall below $0.1$ for any value of $A_1$ are
identified explicitly.  The log likelihood for the source 0629$+$109
is sufficiently low that it is not shown.}
\label{fig:decon}
\end{figure}

\begin{figure}
\caption[Schematic of Outer Galaxy Scattering]
{A schematic of scattering in the outer Galaxy.  Small, dark
regions are molecular clouds from the CO survey of Wouterloot \&
Brand~(1989).  Cross-hatched regions represent turbulent gas,
responsible for the scattering, potentially surrounding these clouds,
resulting from embedded star formation.  This figure illustrates how
the scattering gas could follow the distribution of molecular gas.}
\label{fig:ac.summary}
\end{figure}

\clearpage

\begin{deluxetable}{lccc}
\tablecaption{Scattering Diameters at 1~GHz from the Survey of Lazio \& Cordes~(1997)\label{tab:broaden}}
\tablehead{\colhead{Name} & \colhead{$\ell$} & \colhead{$b$} & 
	\colhead{$\ths$} \\
	& \colhead{(\arcdeg)} & \colhead{(\arcdeg)}
	& \colhead{(mas)}}

\startdata
87GB~0433$+$4706 & 157.3 & $-$0.0  & $<$48\phd\phn \nl
87GB~0451$+$4309 & 162.5 & $-$0.1  & $<$\phn5.1    \nl
87GB~0512$+$2627 & 178.5 & $-$6.6  & $<$\phn6.8    \nl
87GB~0537$+$3059 & 177.9 & \phs0.2 & $<$\phn3.7    \nl
87GB~0547$+$3044 & 178.9 & \phs1.8 & $<$\phn6.1    \nl

\nl

87GB~0558$+$2325 & 186.5 & \phs0.3 &  3.7$\pm$0.8  \nl
87GB~0600$+$2957 & 180.9 & \phs4.0 &  1.5$\pm$0.4  \nl
87GB~0621$+$1219 & 198.8 & $-$0.4  &  2.8$\pm$0.1  \nl
87GB~0622$+$1153 & 199.4 & $-$0.3  &  $<$18        \nl

\enddata
\end{deluxetable}

\begin{deluxetable}{lccccccc}
\tablecaption{Scattering Measurements from the
	Literature\label{tab:literature}}

\tablehead{\colhead{Name} &
	\colhead{$\ell$} & \colhead{$b$} &
	\colhead{$\Delta\nu_{\rm d}$} & \colhead{$\ths$} & 
	\colhead{$D$} & \colhead{Ref.} \\
	& \colhead{(${}\arcdeg$)} & \colhead{(${}\arcdeg$)} & 
	\colhead{(MHz)} & \colhead{(mas)} & \colhead{(kpc)}}

\startdata

PSR~B0301$+$19 & 161.14 & $-$33.27     & \phn9.55$\pm$2.4\phn & 0.29$\pm$0.097 & 0.94   & 1 \nl
PSR~B0320$+$39 & 152.18 & $-$14.33     & \phn2.291$\pm$0.77   & 0.46$\pm$0.15  & 1.47   & 1 \nl
PSR~B0450$+$55 & 152.62 & \phs7.54     & \phn9.55$\pm$2.4\phn & 0.29$\pm$0.097 & 0.78   & 1 \nl
PSR~B0525$+$21 & 183.86 & \phn$-$6.89  & \phn0.62$\pm$0.2\phn & 0.72$\pm$0.24 & 2.27    & 1 \nl
PSR~B0531$+$21 & 184.60 & \phn$-$5.80  & \nodata              & 0.50$\pm$0.05 & 2.0\phn & 4 \nl

\nl

PSR~B0540$+$23 & 184.36 & \phn$-$3.31  & \phn0.11$\pm$0.03    & 1.4\phn$\pm$0.47 & 3.53 & 1 \nl
PSR~B0611$+$22 & 188.79 & \phs\phn2.39 & \phn0.04$\pm$0.01    & 2.0\phn$\pm$0.67 & 4.72 & 1 \nl
PSR~B0626$+$24 & 188.82 & \phs\phn6.22 & \phn0.08$\pm$0.02    & 1.4\phn$\pm$0.47 & 4.69 & 1 \nl
PSR~B0656$+$14 & 201.11 & \phs\phn8.25 & \phn8.51$\pm$2.1\phn & 0.34$\pm$0.11    & 0.76 & 1 \nl
PSR~B0823$+$26 & 196.96 & \phs31.74    & \phn9.55$\pm$2.4\phn & 0.29$\pm$0.097   & 0.37 & 1 \nl

\tableline
\tablebreak

PSR~B1112$+$50 & 154.41 & \phs60.36    & 19.50$\pm$6.4\phn    & 0.26$\pm$0.087   & 0.54 & 1 \nl

\nl

0503$+$467 & 161.00 & \phs\phn3.70 & \nodata   & $<$\phn4 & \nodata & 2 \nl
0629$+$109 & 201.50 & \phs\phn0.50 & \nodata   & 25\phd\phn\phn$\pm$1.4\phn & \nodata & 3 \nl

\enddata
\tablecomments{All quantities have been scaled to 1~GHz.  The
scattering angle for the pulsars depends upon both the scintillation
bandwidth and adopted distance, $\ths =
0.85\,\mathrm{mas}/\sqrt{D_{\mathrm{kpc}}\delnud}$.}

\tablerefs{(1)~Cordes~(1986); (2)~Spangler et al.~(1986); (3)~Dennison
et al.~(1984); (4)~Gwinn et al.~(1993)}
\end{deluxetable}

\begin{deluxetable}{lc}
\tablecaption{Preferred Model Parameters\label{tab:params}}

\tablehead{\colhead{Model} & \colhead{Adopted} \\
	\colhead{Parameter} & \colhead{Value}}

\startdata

radial form & truncated disk \nl

\nl

$n_1$\tablenotemark{a}\ (cm${}^{-3}$) & 0.0188          \nl
$\hin$\tablenotemark{a}\ (kpc)        & 0.88\phn\phn    \nl
$\fin$                                & 0.4\phn\phn\phn \nl

\nl

$f_4$                                & 0.25 \nl

\nl

$A_1$ (kpc)                          & 20\phd\phn    \nl
$\hout$ (kpc)                        & \phn1\phd\phn \nl
$\fout$                              & \phn0.4       \nl
$\Psi$ (\arcdeg)                     & \phn0\phd\phn \nl

\enddata
\tablenotetext{a}{This parameter was not varied in our likelihood
analysis; its value is taken from the Taylor-Cordes model.}

\end{deluxetable}

\begin{deluxetable}{lcccc}
\tablecaption{Radial Extent of Ionized Disks for External
	Galaxies\label{tab:external}}

\tablehead{\colhead{Name} & \colhead{$R$} 
	& \colhead{$L_{\mathrm{FIR}}/D_{25}^2$} &
	\colhead{$\delta(\mathrm{EM})$} & \colhead{Ref.} \\
	& \colhead{(kpc)} 
	& \colhead{$10^{40}$\,erg\,s${}^{-1}$\,kpc${}^{-2}$}
	& \colhead{(pc~cm${}^{-6}$)}}

\startdata
NGC~891  & 15    & \phm{$<$}2.2\phn & \phn6.5 & 1 \nl
NGC~3079 & 11    & \phm{$<$}8.9\phn & 10\phd\phn & 3 \nl
NGC~4013 & \phn9 & \phm{$<$}2.6\phn & \phn2.5 & 4 \nl
NGC~4217 & 12    & $<$0.12          & \phn3.2 & 4 \nl
NGC~4302 & 10    & $<$2.3\phn       & \phn2.1 & 4 \nl

\nl

NGC~4565 & 17    & \phm{$<$}0.3\phn & \phn2.3 & 2 \nl
NGC~4631 & 14    & \phm{$<$}1.8\phn & \phn3.2 & 2 \nl
NGC~4762 & \phn6 & $<$0.15          & \phn2.4 & 4 \nl
NGC~5023 & \phn4 & $<$0.09          & \phn2.3 & 4 \nl
NGC~5746 & 26    & \phm{$<$}0.2\phn & \phn3.7 & 4 \nl

\tableline
\tablebreak

NGC~5907 & 22    & \phm{$<$}0.8\phn & \phn3.2 & 4 \nl
UGC~4278 & \phn8 & $<$0.04          & \phn2.7 & 4 \nl
UGC~10288 & 16   & \phm{$<$}0.4\phn & \phn4.4 & 4 \nl

\nl

Galaxy   & 10    & \phm{$<$}3.0\phn & \nodata & 5 \nl

\enddata

\tablerefs{(1)~Rand, Kulkarni, \& Hester~(1990); (2)~Rand, Kulkarni,
\& Hester~(1992); (3)~Veilleux,
Cecil, \& Bland-Hawthorn~(1995); (4)~Rand~(1996); (5)~this work}
\end{deluxetable}

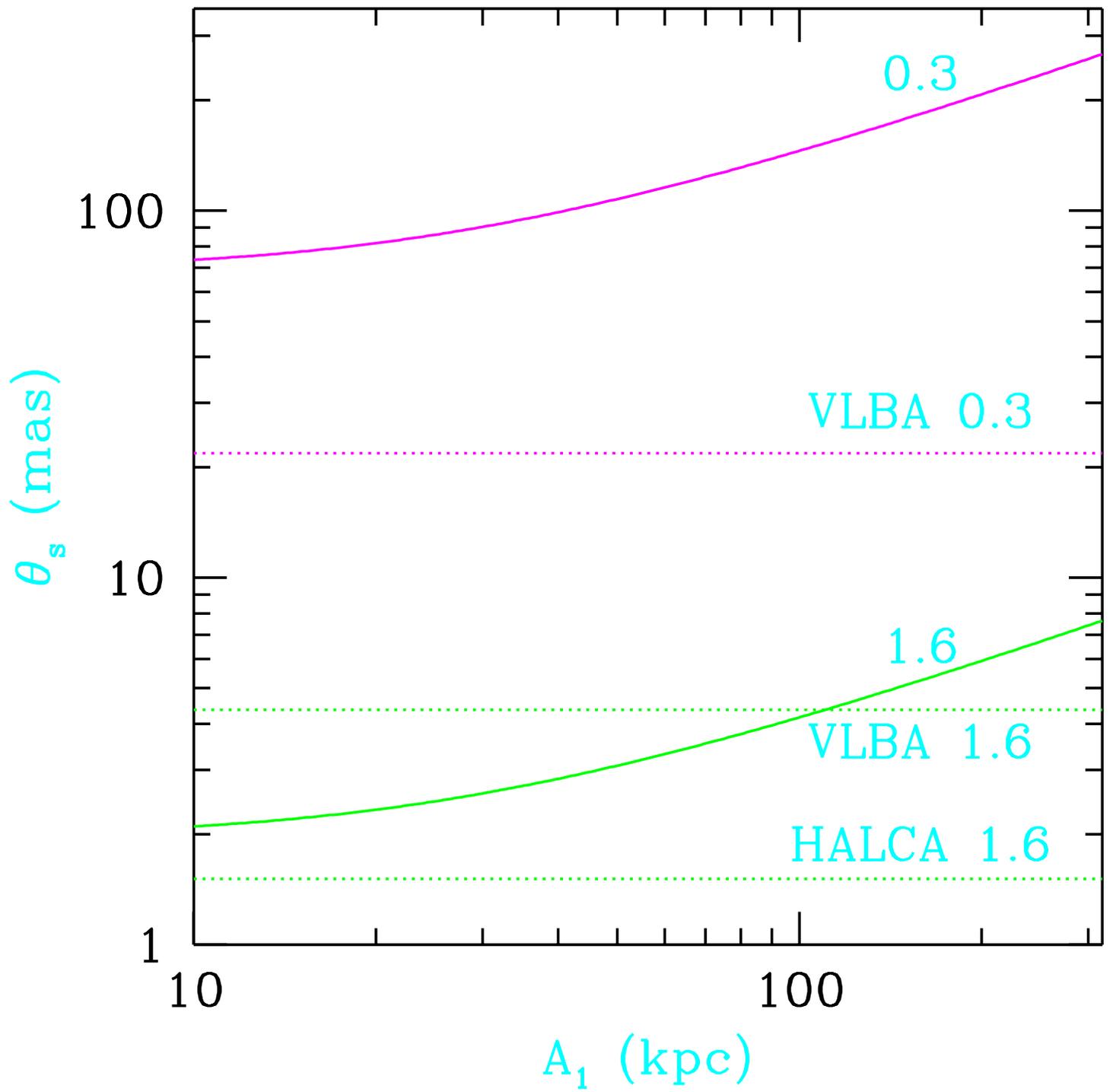
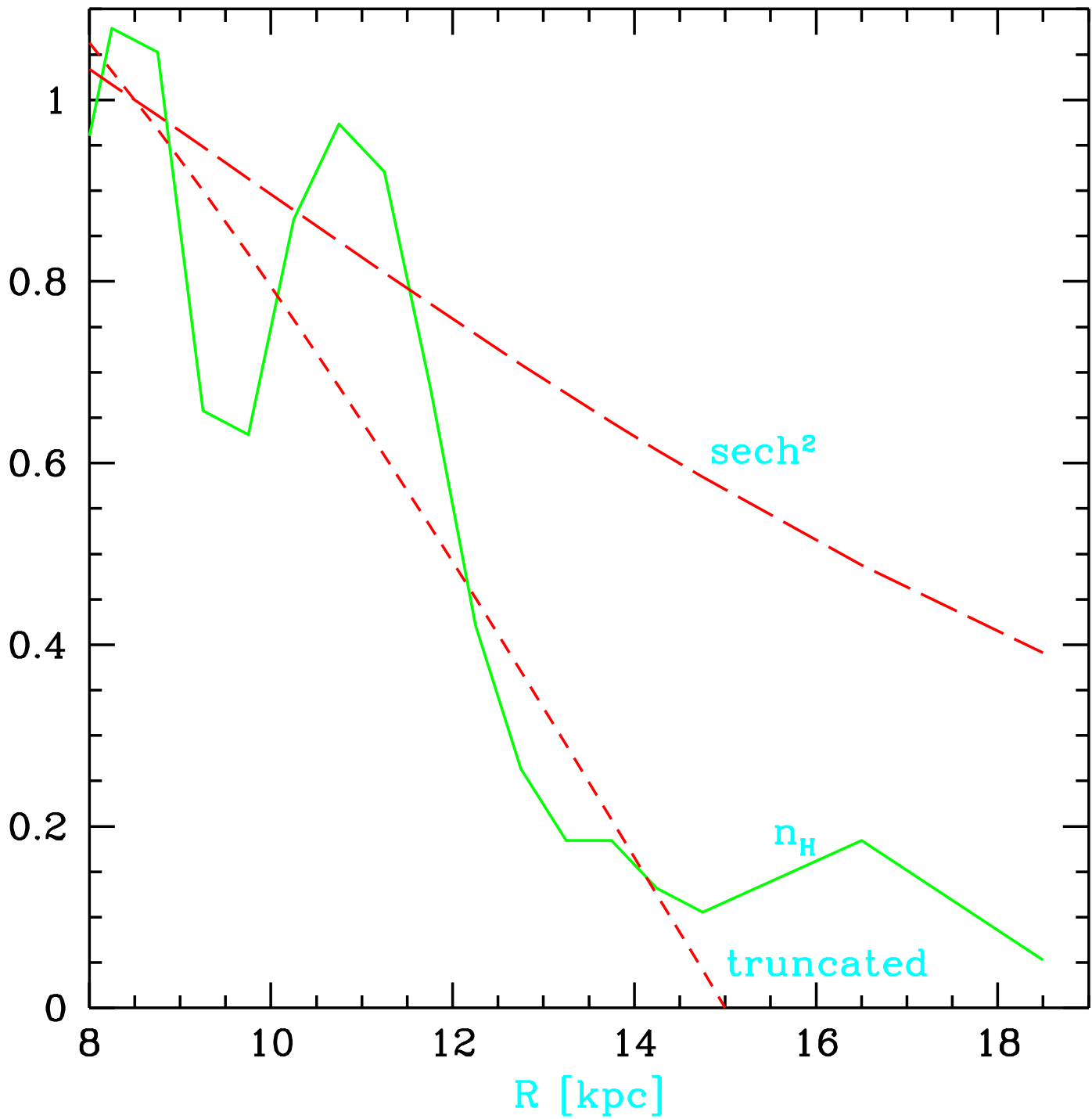
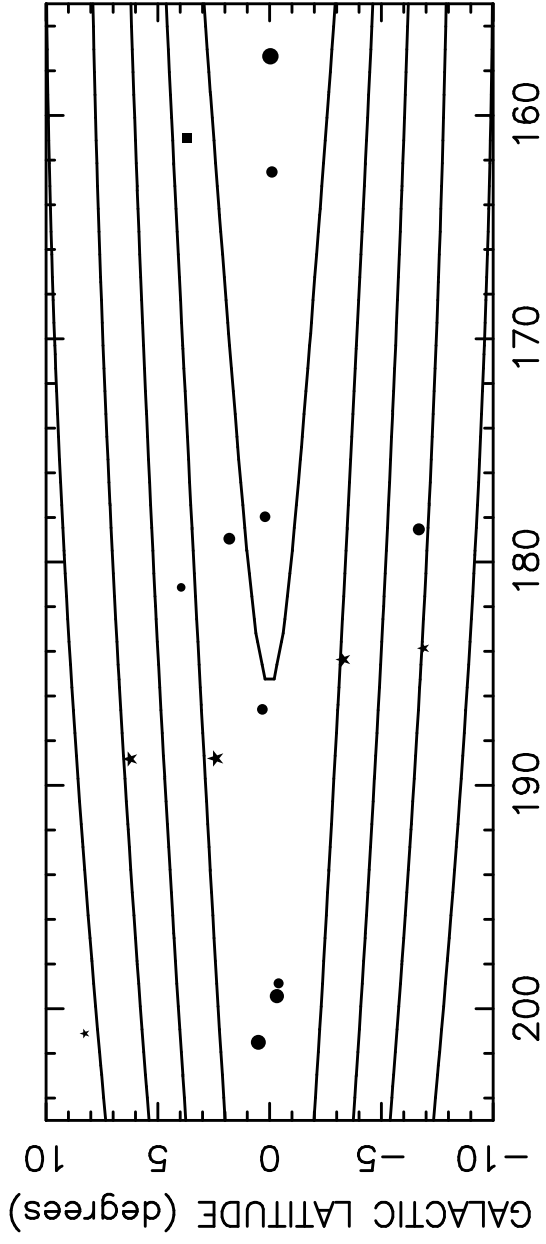
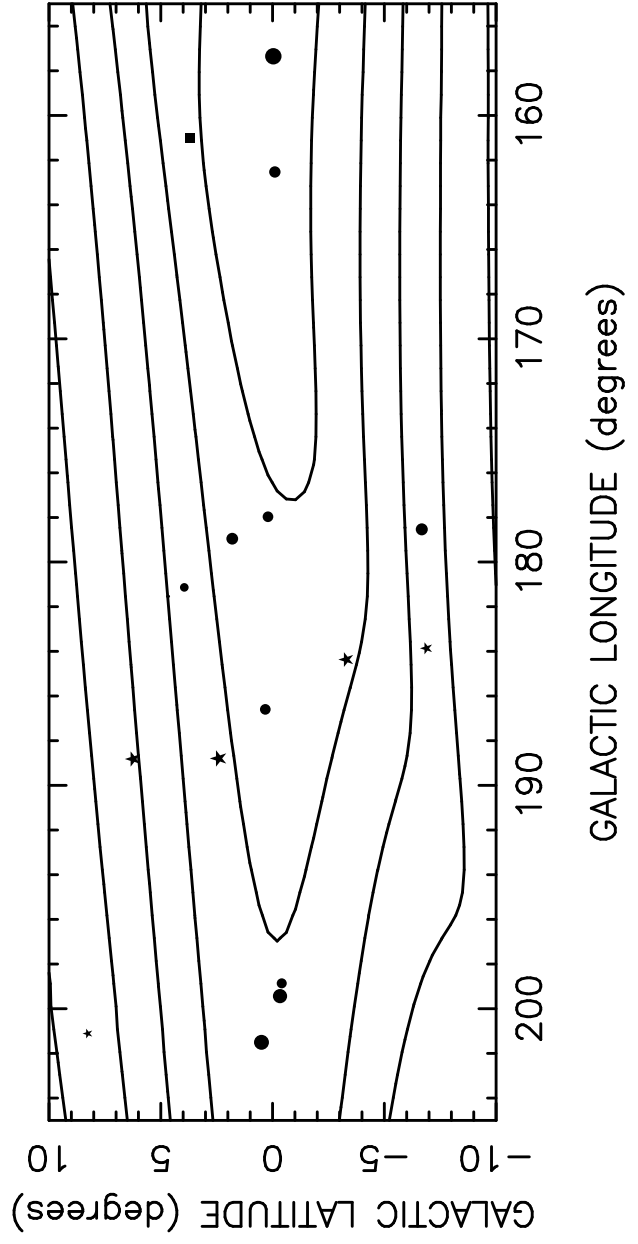
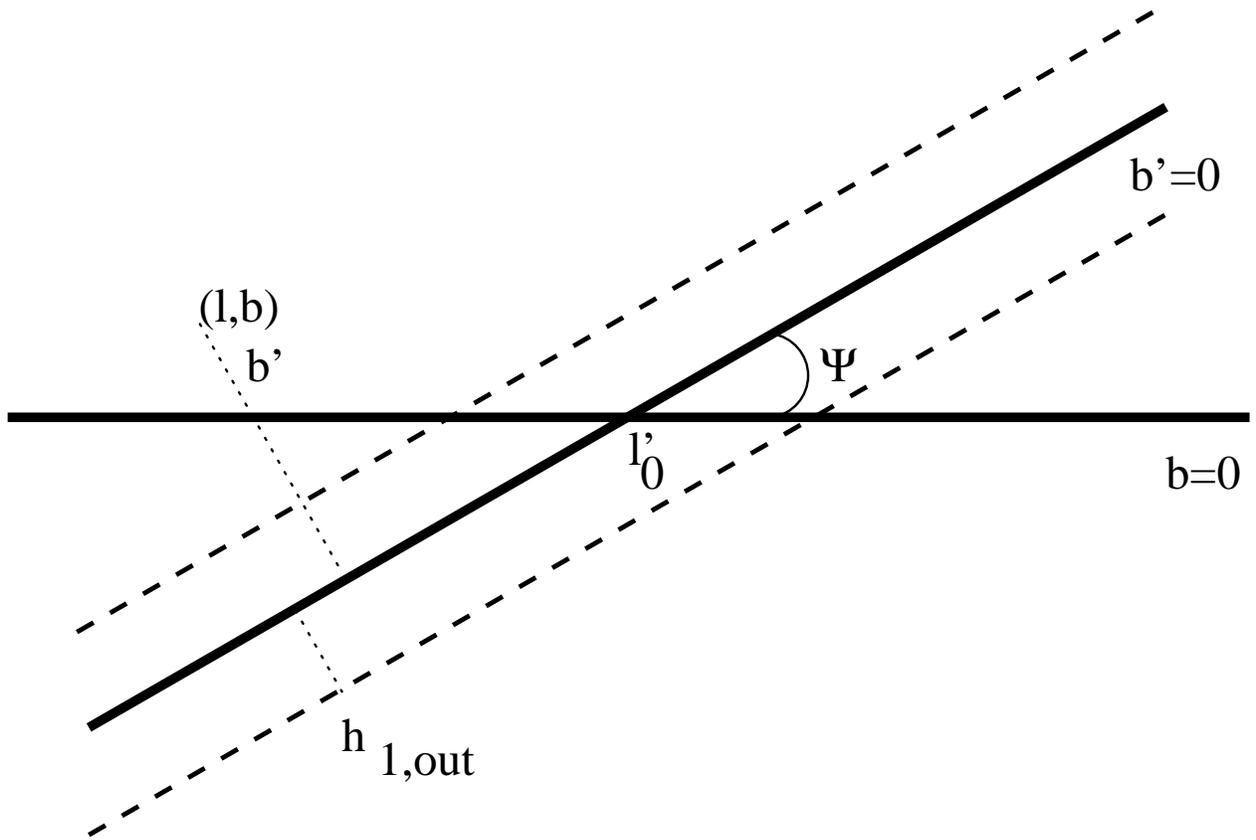
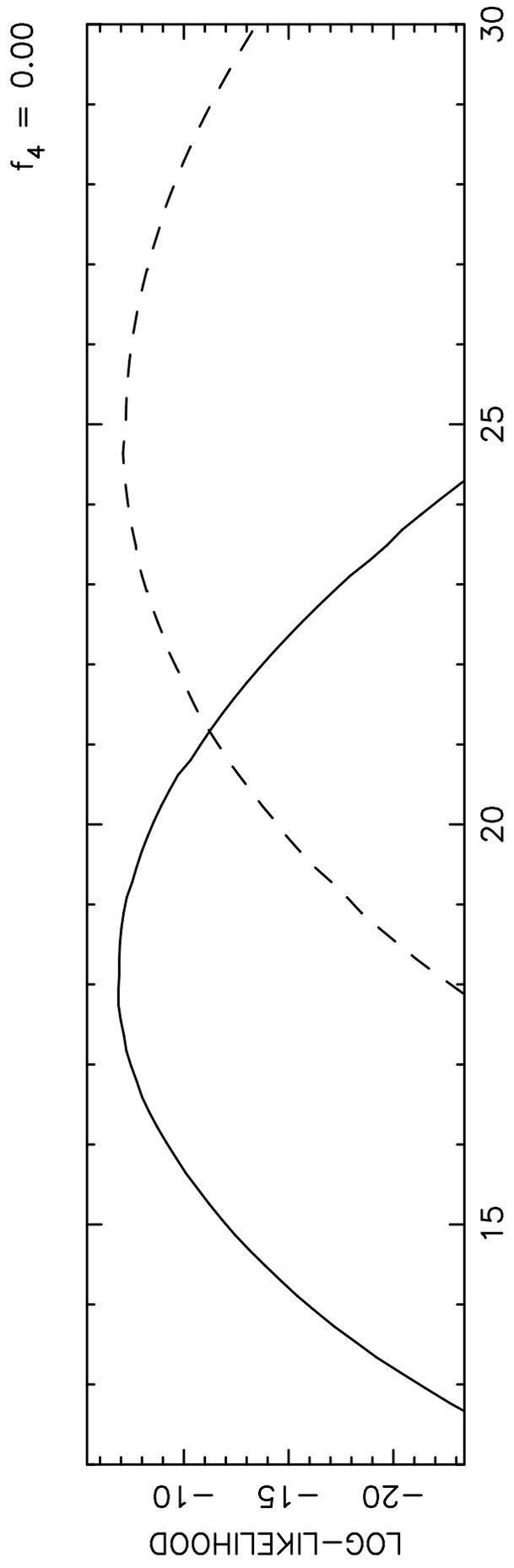
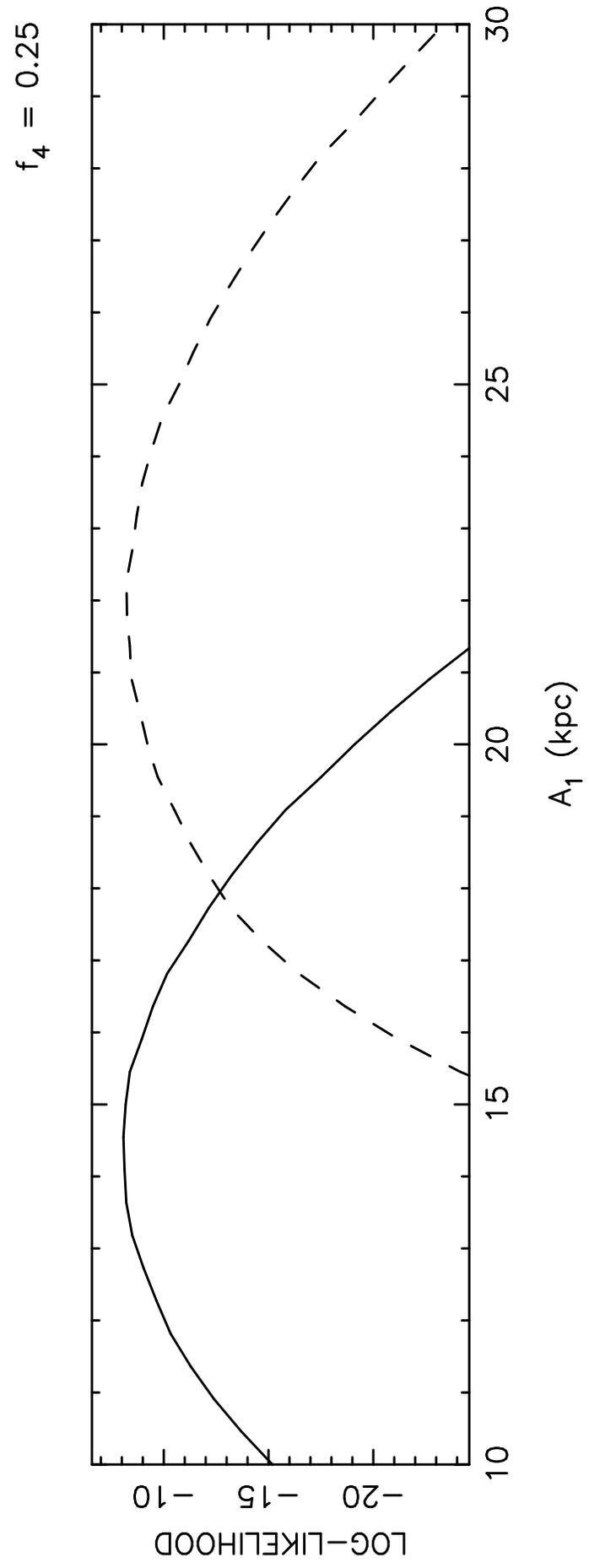
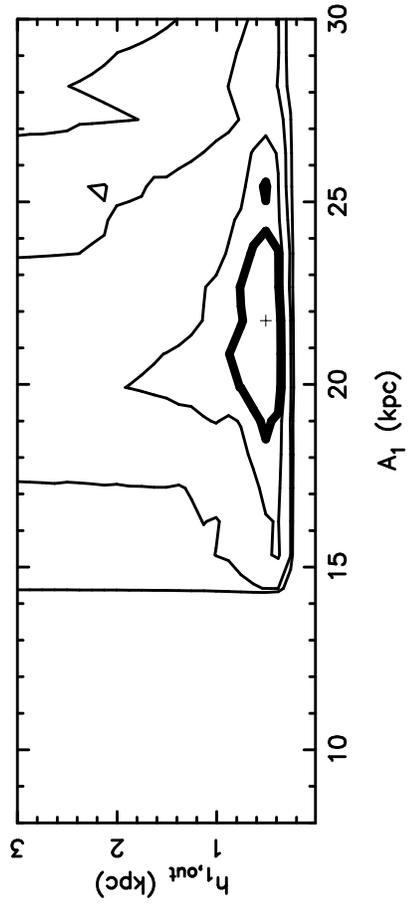
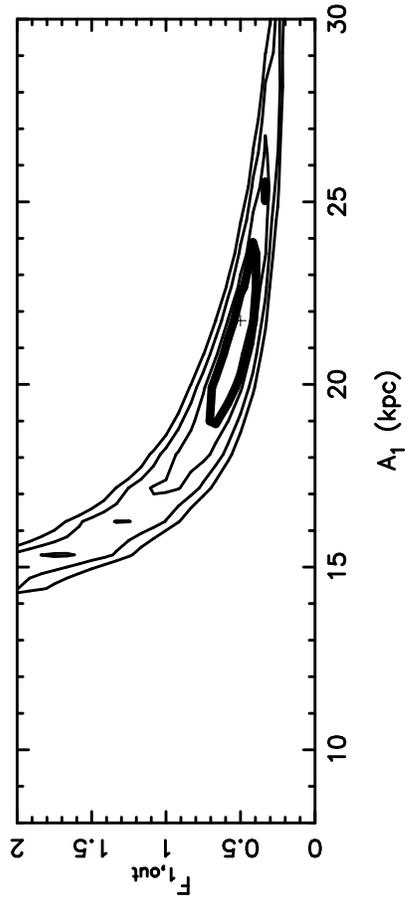
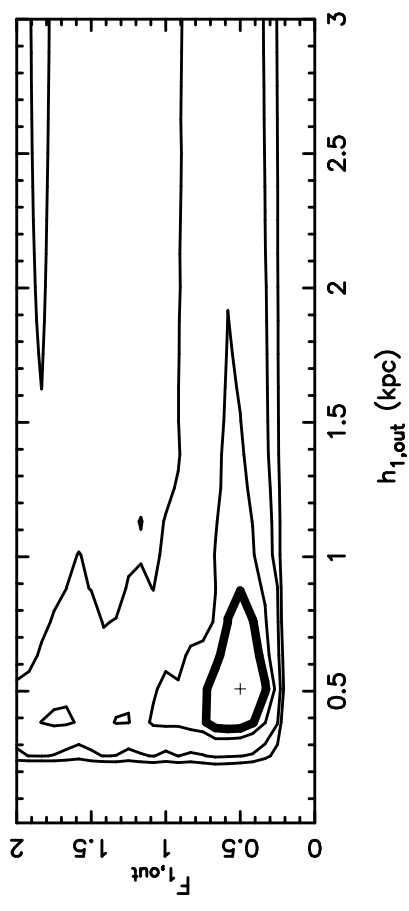
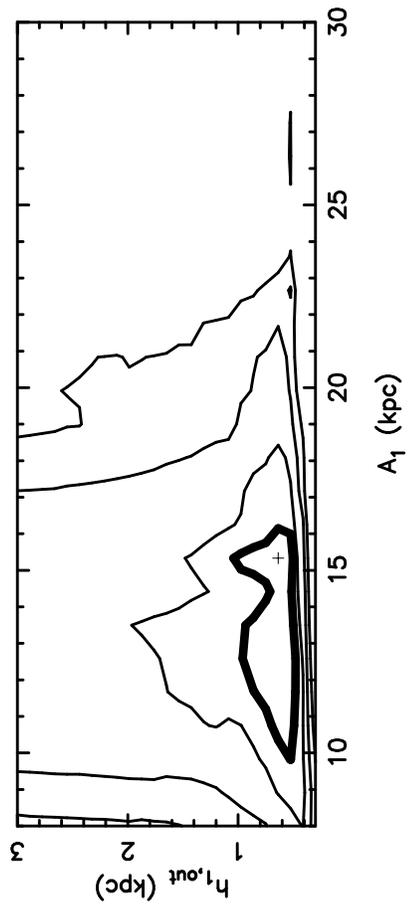
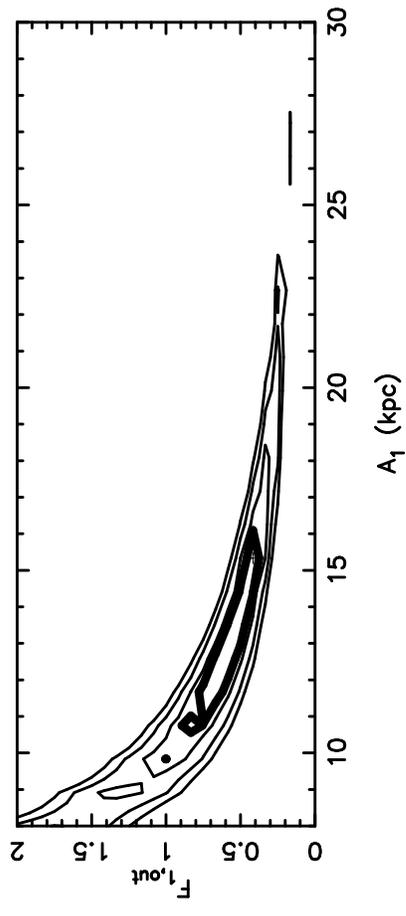
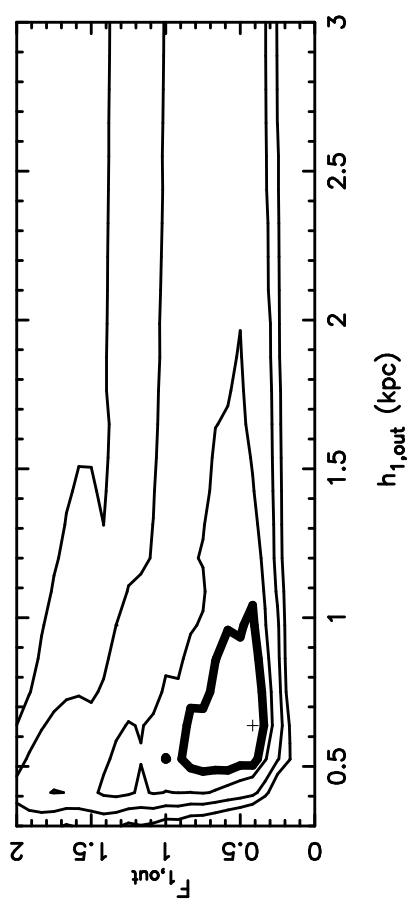
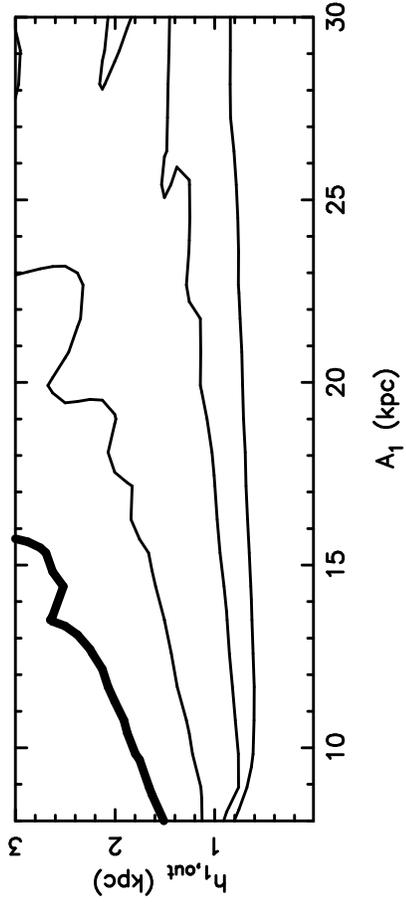
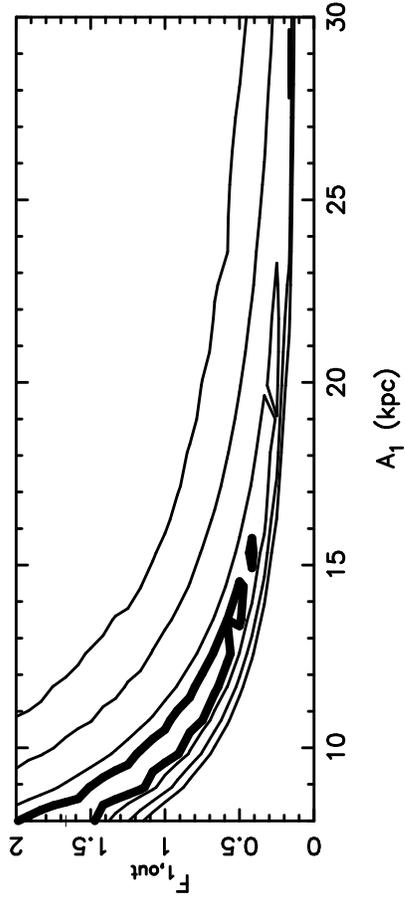
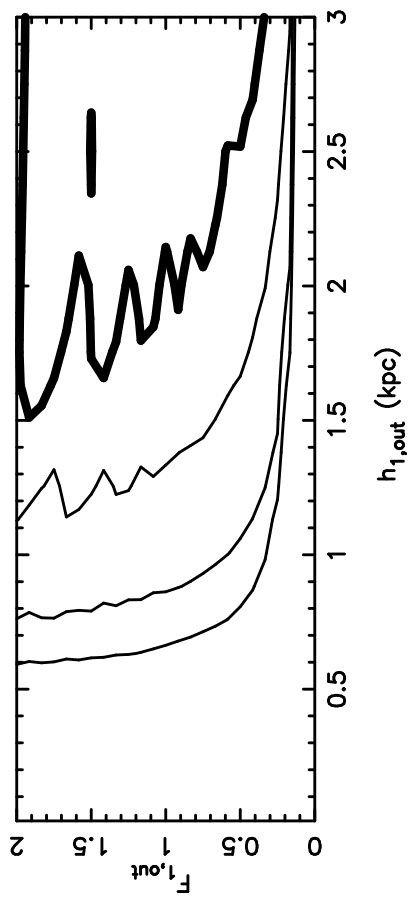
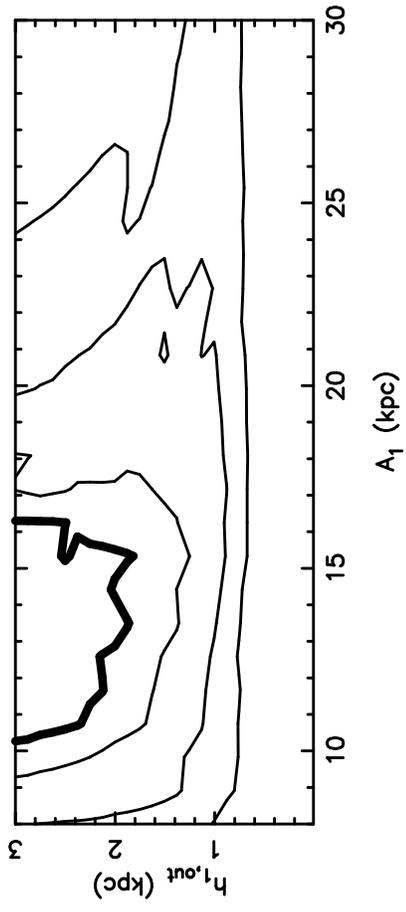
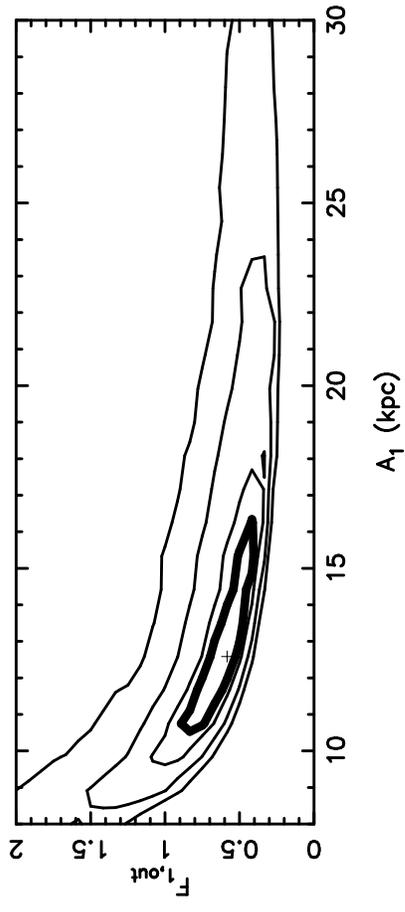
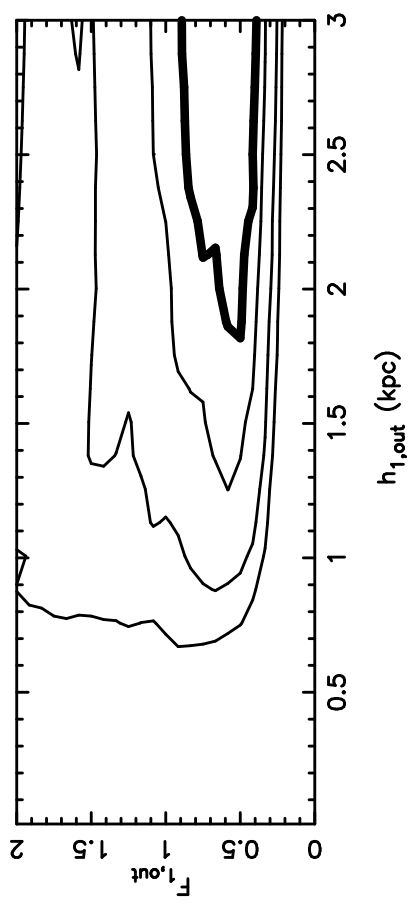
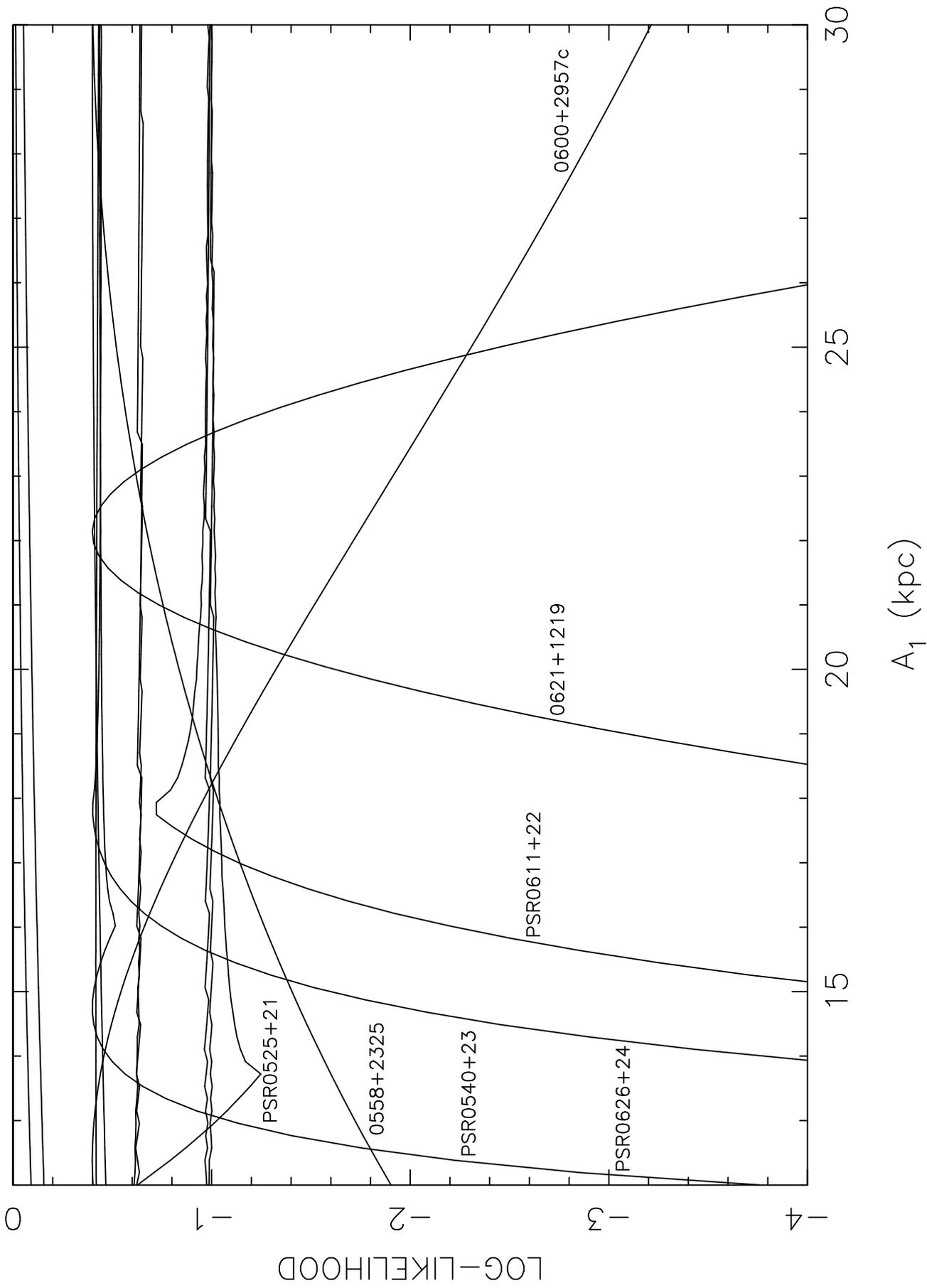
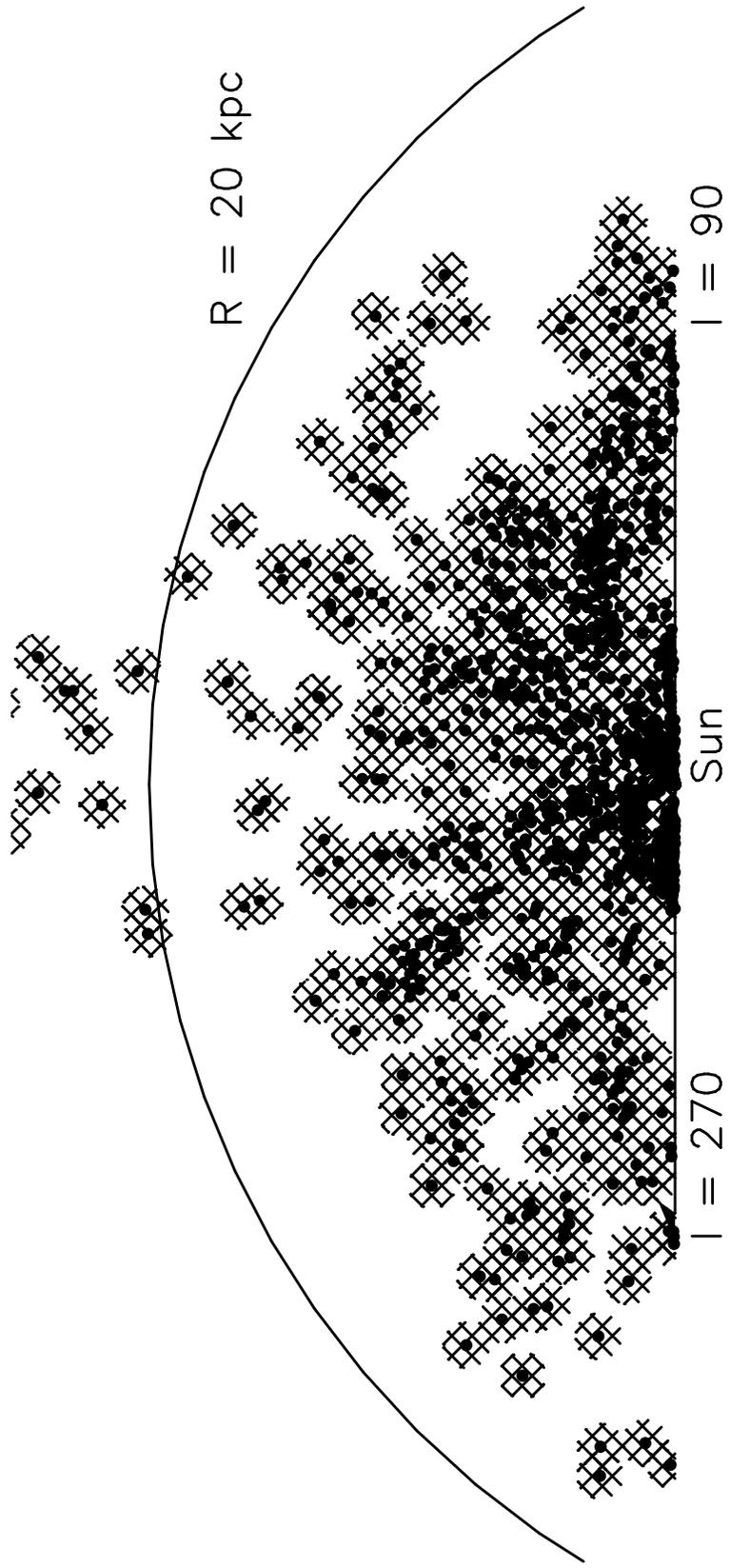
\end{document}